%% file: main.tex
\documentclass[journal=jacsat,manuscript=article]{achemso}
\usepackage[version=3]{mhchem} % Formula subscripts using \ce{}
\usepackage{amsmath,amsthm,amssymb,amsfonts}
\usepackage{siunitx}
\usepackage{graphicx}
\usepackage{booktabs}
\usepackage[english]{babel}
\usepackage{textgreek}
\usepackage{hyperref}
\usepackage{xcolor}
\usepackage{tikz}
\usetikzlibrary{arrows.meta, positioning, fit, backgrounds, calc, shapes.geometric}

\DeclareSIUnit\angstrom{\text {Å}}

\author{Debarshi Banerjee}
\affiliation{International Centre for Theoretical Physics (ICTP), Strada Costiera 11, 34151 Trieste, Italy}
\alsoaffiliation{Scuola Internazionale Superiore di Studi Avanzati (SISSA), via Bonomea 265, 34136 Trieste, Italy}

\author{Gonzalo Díaz Mirón}
\affiliation{International Centre for Theoretical Physics (ICTP), Strada Costiera 11, 34151 Trieste, Italy}

\author{Alex Rodriguez}
\affiliation{International Centre for Theoretical Physics (ICTP), Strada Costiera 11, 34151 Trieste, Italy}
\alsoaffiliation{Dipartimento di Matematica, Informatica e Geoscienze, Universitá degli Studi di Trieste, via Alfonso Valerio 12/1, 34127, Trieste, Italy}

\author{Ali Hassanali}%
\email{ahassana@ictp.it}
\affiliation{International Centre for Theoretical Physics (ICTP), Strada Costiera 11, 34151 Trieste, Italy}

\title{Machine learning the non-radiative decay modes in photochemical processes}

\abbreviations{ML,TSH,NAD}

\SectionNumbersOn

\begin{document}

\begin{abstract}

Non-radiative decay in photoexcited molecular systems is driven by nuclear motion that guides the system toward conical intersections (CIs), where electronic states become degenerate and nonadiabatic transitions occur. Identifying the nuclear degrees of freedom responsible for accessing these regions from nonadiabatic molecular dynamics (NAD) simulations remains challenging due to the high dimensionality and collective nature of the underlying motions. Here, we introduce an unsupervised, information-theoretic framework based on the Differentiable Information Imbalance (DII) to identify the nuclear coordinates governing CI access directly from trajectory surface hopping (TSH) simulations. By quantifying correlations between structural descriptors and electronic observables, including energy gaps and oscillator strengths, the method ranks nuclear degrees of freedom according to their predictive relevance. A multi-step analysis protocol further enables the extraction of low-dimensional, physically interpretable modes associated with non-radiative decay. We apply the framework to a diverse set of systems, including the methaniminium cation, furan, L-glutamine, L-pyroglutamine–ammonium, and a photoactive molecular motor. Across all systems, the method recovers known mechanistic coordinates governing non-radiative decay while also providing insight into the relative importance of competing modes when multiple structural distortions contribute to CI access. The analysis additionally reveals a systematic distinction between observables: energy gaps are typically controlled by a small number of localized coordinates, whereas oscillator strengths depend on more collective and distributed structural rearrangements. Overall, the DII-based framework combines predictive power with direct interpretability, providing a general and scalable route for extracting mechanistic insight from high-dimensional data coming from NAD simulations and constructing reduced-dimensional models of excited-state dynamics.
\end{abstract}

%%%%%%%%%%%%%%%%%%%%%%%%
\section{Introduction}\label{sec:intro}

Photoinduced processes in molecular systems are frequently governed by chemistry associated with ultrafast transitions between electronic states. In polyatomic molecules, these transitions are typically mediated by conical intersections (CIs)\cite{Yarkony1996,Levine2007,Matsika2021}, which represent points of degeneracy between adiabatic potential energy surfaces. At such geometries, the Born–Oppenheimer approximation breaks down and nuclear and electronic motions become strongly coupled, enabling efficient non-radiative relaxation pathways. Conical intersections therefore play a central role in photochemical reactivity, controlling excited-state lifetimes and branching ratios in systems ranging from small organic chromophores to biological molecules and photoactive materials.

A widely used theoretical framework for investigating these processes is trajectory surface hopping (TSH), which allows for simulating non-adiabatic molecular dynamics (NAD) in high-dimensional systems in a computationally efficient manner. Modern implementations allow the generation of large ensembles of excited-state trajectories that capture complex relaxation mechanisms and competing decay channels. Despite this progress, extracting mechanistic insight from such simulations remains challenging.\cite{Dreuw2026} In particular, it is often unclear which nuclear degrees of freedom actively drive the approach to the conical intersection within the high-dimensional configuration space sampled by the trajectories.

From a theoretical perspective, the topology of a conical intersection is locally characterized by the branching space, spanned by the gradient difference and non-adiabatic coupling vectors (NACVs), which lift the degeneracy between the intersecting electronic states.\cite{Atchity1991,Yarkony2001,Domcke2012,Gozem2014} However, this description is strictly local and typically obtained at optimized minimum-energy conical intersection (MECI) geometries. In dynamical simulations, trajectories may approach the intersection along collective nuclear distortions that are not trivially related to these local vectors, and the relevant coordinates may involve nonlinear combinations of atomic motions distributed across the molecule.\cite{Barbatti2006,Virshup2012} As a consequence, identifying the dominant degrees of freedom governing access to CIs from a large ensemble of trajectories remains a nontrivial problem. This challenge becomes even more pronounced for larger systems where the number of nuclear degrees of freedom grows rapidly and the relevant motions are often distributed across many coupled coordinates.\cite{Li2017,Li2018,Mai2019,Zhu2022,Zhu2024}

In recent years, machine learning techniques have become increasingly embedded in both the execution and analysis of molecular simulations.\cite{jackson2023,Westermayr2021} The behavior of molecular systems is generally governed by fluctuations across high-dimensional free-energy landscapes,\cite{laio2021} which complicates the task of extracting clear mechanistic interpretations from simulation data. This challenge has motivated substantial efforts toward developing data-driven methodologies capable of characterizing, interrogating, and rationalizing the complex dynamical behavior observed in molecular simulations.
%The identification of a reduced set of collective variables (CVs) that capture the essential nuclear motions governing complex molecular processes is a long-standing challenge in molecular sciences, underpinning tasks ranging from free energy calculations to reaction mechanism elucidation. 
In the context of nonadiabatic molecular dynamics (NAD), the selection of such coordinates has often relied on chemical intuition and visual inspection of trajectories, where key distortions are inferred by examining representative geometries or displacement patterns associated with electronic transitions.

A variety of different approaches have been proposed to identify the nuclear degrees of freedom governing excited-state dynamics and nonradiative decay. Early efforts largely relied on normal mode analysis or principal component analysis (PCA) to extract dominant motions from molecular dynamics trajectories, which were then used to construct reduced-dimensionality models such as linear vibronic coupling (LVC) Hamiltonians or to guide quantum dynamics simulations \cite{Zauleck2016, Capano2017, Mai2019, Delmas2025}. However, normal mode analysis is intrinsically a local, harmonic description defined around a reference minimum and is therefore most appropriate for small-amplitude motions in relatively rigid systems. In strongly anharmonic regimes, such as those involving large structural rearrangements or dissociation, the normal mode picture breaks down and may fail to capture the relevant nuclear distortions driving the dynamics. PCA has also been widely applied to trajectory surface hopping simulations to identify collective nuclear motions associated with internal conversion or intersystem crossing pathways \cite{Capano2017, Atkins2017, Peng2021, Zhu2022, Rukin2024}. However, because PCA ranks modes according to structural variance, it may emphasize large-amplitude or low-frequency motions that are not directly relevant to nonadiabatic transitions, while neglecting smaller-amplitude motions that lead to electronic transitions. This limitation has been explicitly observed in several systems, where dominant principal components correspond to collective motions which may not necessarily lead to non-adiabatic transitions.\cite{Mai2019}

To overcome these limitations, nonlinear dimensionality reduction techniques, such as diffusion maps and related manifold learning approaches, have been employed to extract low-dimensional reaction coordinates directly from nonadiabatic dynamics simulations \cite{Virshup2012,Belyaev2015,Richings2021}. While these methods can capture nonlinear correlations in the data, the resulting collective variables are often difficult to interpret physically and require additional, sometimes ad hoc, procedures to map them back onto chemically meaningful internal coordinates \cite{Li2017, Li2018}. Alternative strategies based on clustering combined with PCA or multidimensional scaling have also been used to identify distinct dynamical channels, with mechanistic insight obtained by comparing representative structures across clusters \cite{Li2018, Zhu2022}. In parallel, information-theoretic approaches have been introduced to quantify the relationship between structural descriptors and electronic observables, for example by computing mutual information between internal coordinates or atom-centered features and quantities such as energy gaps or nonadiabatic couplings \cite{Tavadze2018, How2021, How2022, Zhou2020, Mangan2021}. While these methods provide a more direct link between nuclear motion and electronic response, they typically operate at the level of individual descriptors or atoms, and do not directly yield collective modes suitable for reduced-dimensionality dynamical models. These considerations highlight the need for approaches that can identify collective nuclear coordinates directly from dynamical data, while explicitly incorporating their relevance to electronic structure changes and nonadiabatic processes.

In this work we address this problem using the information imbalance framework introduced by Glielmo and co-workers\cite{glielmo2022ranking, wild2025automatic}, which provides an information-theoretic approach to quantify the predictive relationship between different representations of a system. The method evaluates how well the neighborhood structure defined by one set of variables can reproduce that defined by another, thereby enabling the identification of coordinates that retain the most relevant information about a target observable.\cite{Donkor2023,Donkor2024,DiPino2025} In the context of non-adiabatic dynamics, this allows us to determine which nuclear degrees of freedom best encode the approach to the conical intersection as reflected in quantities such as the electronic energy gap or hopping events along trajectory surface hopping simulations.

By applying this framework to ensembles of TSH trajectories, we identify low-dimensional collective coordinates governing CI access directly from dynamical data, without relying on optimized intersection geometries or predefined reaction coordinates. We first demonstrate the approach on well-characterized model systems, including the methaniminium cation\cite{Barbatti2006,Barbatti2007,Hollas2018,Suchan2020} and furan\cite{Fuji2010,Oesterling2017}, where the extracted motions can be directly compared with established mechanistic interpretations. We then extend the analysis to two amino-acid systems, L-glutamine and L-pyroglutamine-ammonium, which are central to the phenomenon of non-aromatic fluorescence studied extensively in our group (NAF)\cite{arnon2021off,kumar2022role,stephens2021short,pinotsi2016proton,Grisanti_JACS_2020,morzan_2022_jpcb,CO_lock_2023,Banerjee2025,Monti2025} and exhibit more complex photophysical behavior. NAF refers to photoinduced light emission from chemical systems that lack conventional aromatic or extended $\pi$-conjugated chromophores\cite{new_chen2018prevalent,new_tomalia2019non,new_tang2021nonconventional,morzan_2022_jpcb,zhang2025_gluser}. In the specific cases considered here, experiments showed that L-glutamine is not optically active, but upon heating in water and complexation with an ammonium ion it undergoes a chemical transformation to form L-pyroglutamine-ammonium, a supramolecular assembly stabilized by short hydrogen bonds and displaying enhanced fluorescence.\cite{stephens2021short} Finally, we consider a well-studied class of photoactive molecular motors\cite{Conyard2012,Kazaryan2011,Garcia2020,Pang2017,Wen2023}, illustrating the ability of the method to uncover mechanistically relevant degrees of freedom in larger and structurally diverse systems.

The paper is organized as follows. Section~2 introduces the Information Imbalance framework, and the analysis protocol used to identify nuclear coordinates relevant to non-radiative decay pathways. Section~3 presents applications to the methaniminium cation, furan, L-glutamine, L-pyroglutamine-ammonium, and a photoactive molecular motor, focusing on the coordinates controlling energy gaps and oscillator strengths. Finally, Section~4 summarizes the main conclusions and discusses the implications of the framework for extracting interpretable reduced-dimensional models of excited-state dynamics.

%The proposed framework provides a systematic and data-driven route to extract mechanistic insight from NAD simulations. By identifying the minimal set of nuclear coordinates that control access to conical intersections, the approach facilitates the construction of reduced-dimensional models of excited-state dynamics and offers a general strategy for interpreting large ensembles of trajectory surface hopping simulations.

\begin{figure}[H]
  \centering
  \includegraphics[width=1.0\linewidth]{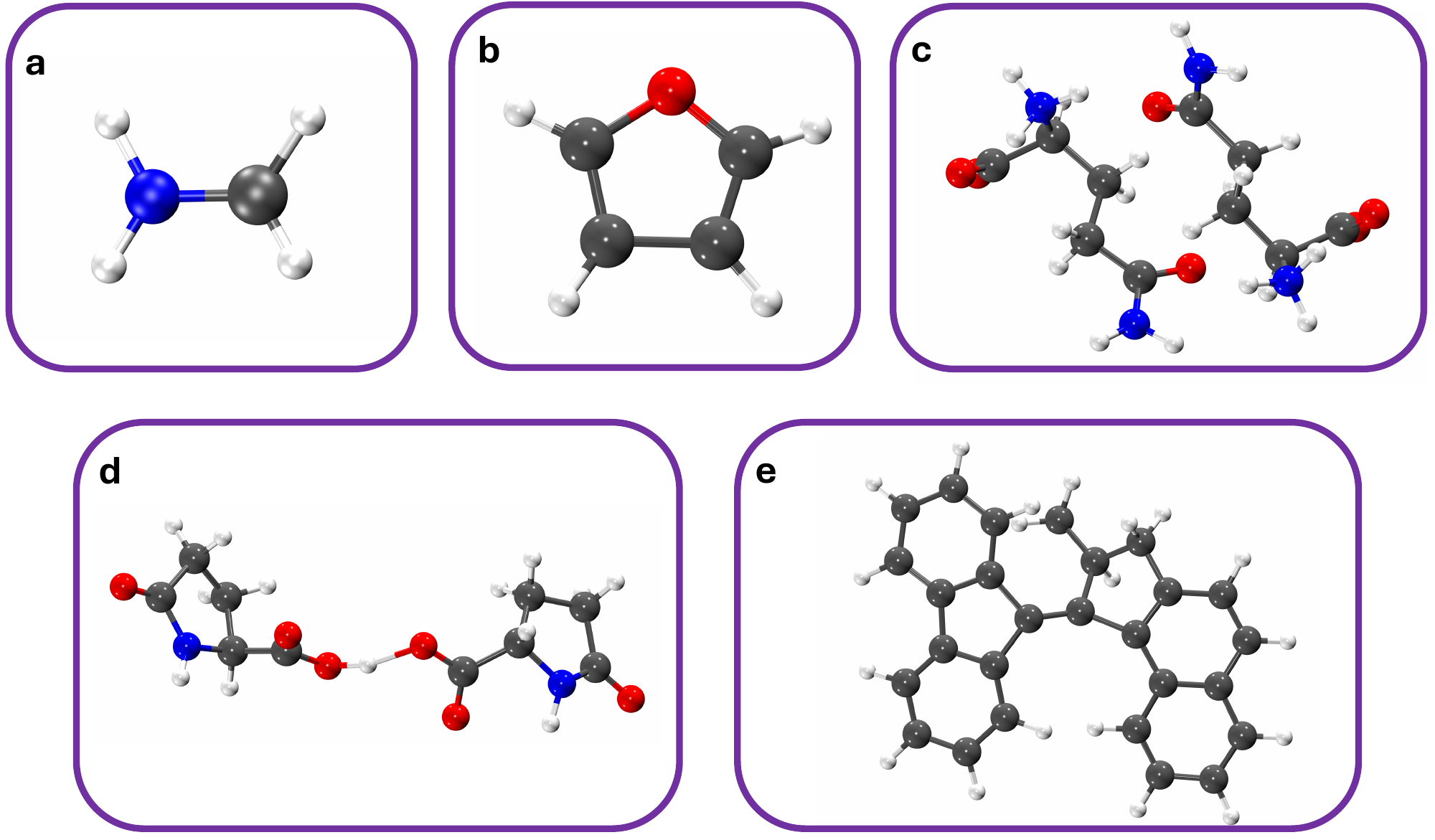}
\caption{
Molecular systems investigated in this work: (a) Methaniminium cation, (b) Furan,
(c) L-glutamine, (d) L-pyroglutamine–ammonium, and (e) Molecular Motor: 9-(2,4,7-trimethyl-2,3-dihydro-1H-inden-1-ylidene)-9H-fluorene.
}  
\label{fig:systems}
\end{figure}

%%%%%%%%%%%%%%%%%%%%%%%%
\section{Methods}\label{sec:methods}
%%%%%%%%%%%%%%%%%%%%%%%%

In this work, we employ an unsupervised machine learning approach based on the Information Imbalance (II) method to identify the nuclear degrees of freedom most relevant for understanding the mechanism associated with non-radiative decay. The II framework, the NAD dataset, and the analysis protocol are described next.

\subsection{Information Imbalance (II)}

The \emph{information imbalance} (II) is a statistical test, introduced by Glielmo \emph{et al.},\cite{glielmo2022ranking} which quantifies the relationship between different distance measures defined over the same dataset. Within the present context, we use II to determine whether structural coordinates can explain variations in observables such as the electronic energy gap between states. Since vanishing energy gaps correspond to the approach to a conical intersection, identifying structural variables that encode the energy-gap landscape provides a direct route to identifying the nuclear motions responsible for accessing the crossing of two potential energy surfaces. In cases where oscillator strengths are also available, the same framework can be used to identify structural coordinates that correlate with the emergence of bright and dark states, corresponding to regions of high and low oscillator strength, respectively. We therefore begin by briefly reviewing the information imbalance formalism.

Information imbalance provides a quantitative measure for comparing the information content of two distance metrics, $d_A$ and $d_B$, defined over the same set of data points. In our application, $d_B$ corresponds to distances defined in the space of electronic observables (such as the energy gap or oscillator strength), while $d_A$ corresponds to distances defined in a structural representation of the system, such as internal coordinates or Coulomb matrix elements (which will be introduced in the next section).

For example, distances in the electronic observable space can be defined between two configurations $i$ and $j$ as $d_B^{ij} = \left\| \Delta E^{i} - \Delta E^{j} \right\|, $
where $\Delta E^{i}$ and $\Delta E^{j}$ denote the electronic energy gaps associated with configurations $i$ and $j$, respectively, and $\|\cdot\|$ denotes the Euclidean norm. Distances $d_A^{ij}$ are defined analogously using a chosen set of structural coordinates, or Coulomb Matrix elements.

Given a distance metric $d_A$, one can define $r_A^{ij}$ as the \emph{rank} of point $j$ among all points in the dataset when ordered by their distance from point $i$ according to $d_A$. Similarly, $r_B^{ij}$ denotes the rank of point $j$ with respect to point $i$ according to the distance metric $d_B$. Intuitively, the representation $A$ is considered informative about representation $B$ if configurations that are close neighbors according to $d_A$ also remain close neighbors according to $d_B$. In other words, $d_A$ is informative with respect to $d_B$ when the neighborhood structure defined by $d_A$ preserves the neighborhood structure defined by $d_B$.

The II from $d_A$ to $d_B$ is defined as:

\begin{equation}\label{eq:iib}
  \Delta(d_A \rightarrow d_B) := \frac{2}{N} \langle r_B \mid r_A\, \leq k \rangle,
\end{equation}

where $N$ is the total number of points in the data set and $k$ is the number of nearest neighbor points that are considered. Eq.\ref{eq:iib} gives us the II, represented as $\Delta(d_A \rightarrow d_B)$, as a number between 0 and 1. $\Delta(d_A \rightarrow d_B) =0$ occurs when all points that are nearest neighbors according to $d_A$ remain nearest neighbors in $d_B$, whereas $\Delta(d_A \rightarrow d_B) =1$ occurs when the nearest neighbors according to $d_A$ are randomly distributed in $d_B$. The former case is when $d_A$ is maximally informative with respect to $d_B$, and the latter is when it is minimally informative.

% Donkor2023 -> lowest value = 0.3 (order params->SOAP), 0.15/0.2 (SOAP->order params)
% Donkor2024 -> lowest value = 0.3 (this is a messy result but whatever)
% Wild2024 -> not a molecular system; lowest value = 0.7
% wild2025automatic -> lowest value for the FES of chignolin = 0.18 (subset of str features -> all atom RMSD)
% DiPino2025 -> lowest value = 0.4/0.45 (structural features -> electric field)

A recent development extends the original II framework by introducing a differentiable formulation.\cite{wild2025automatic} In this approach, the individual features that define a distance metric are multiplied by learnable weights, which are optimized through gradient-based methods. By allowing the weights to adapt during optimization, this formulation naturally accommodates heterogeneous features that may differ substantially in their physical units or numerical scales. Consequently, it avoids the need for manual pre-processing or ad hoc rescaling of the input variables.

This approach, termed the \emph{Differentiable Information Imbalance} (DII), enables the automatic identification of the most informative subset of structural descriptors for a given dataset. By learning the optimal weighting of features that best preserves the neighborhood structure of a target representation, DII effectively determines which coordinates contribute most strongly to defining the relevant distance metric. The DII can be formally written as

\begin{equation}\label{eq:dii}
  DII(d_A(\pmb{w}) \rightarrow d_B) := \frac{2}{N^2} \sum_{i,j=1}^{N} c_{ij} (\lambda, d_A(\pmb{w}))r^{ij}_B
\end{equation}

where,
\begin{equation}
  c_{ij}(\lambda , d_A(\pmb{w})) := \frac{e^{-d^{ij}_A(\pmb{w})/\lambda}}{\sum_{m\neq i} e^{-d_A^{im}(\pmb w)/\lambda}}
\end{equation}

%Here, $\pmb{w}$ are variational parameters to be optimized via gradient descent in order to assign different weights to different features in distance $d_A$, and the parameter $\lambda$ is chosen according to the average and minimum nearest neighbor distances. In this way, it is possible to both automatically identify the most relevant features in distance $d_A$ that are maximally predictive with respect to distance $d_B$, as well as the respective ratios in which they should be combined. In the limit $\lambda \rightarrow 0$, the DII (Eq. \ref{eq:dii}) is equivalent to and can be viewed as a continuous version of the standard Information Imbalance (Eq. \ref{eq:iib}).

%In this work, we primarily applied the DII in a series of steps that reduce the dimensionality of our problem, to determine the information content of a set of internal degrees of freedom features on the energy gaps along the TSH trajectories. To perform a feature search and to avoid the problem of combinatorial explosion that would naturally arise if one were to test every possible feature subset, we use the forward greedy search algorithm introduced in Ref.\citenum{glielmo2022ranking} coupled with the DII approach that is implemented in the Python package DADApy \cite{glielmo2022dadapy}.

In this expression, $\pmb{w}$ denotes a set of variational parameters that are optimized via gradient descent to assign different relative weights to the features defining the distance $d_A$. The hyperparameter $\lambda$ is selected based on characteristic length scales of the dataset, typically determined from the average and minimum nearest-neighbor distances. Through this optimization procedure, the method simultaneously identifies which features in $d_A$ contain the most predictive information with respect to $d_B$ and determines the optimal combination in which these features should contribute to the distance metric. In the limiting case $\lambda \rightarrow 0$, the DII formulation (Eq.~\ref{eq:dii}) reduces to a continuous analogue of the standard Information Imbalance defined in Eq.~\ref{eq:iib}.

In realistic applications where Information Imbalance (II) has been employed across diverse problems, there is no universal threshold for what constitutes an informative value; however, empirical benchmarks can be established. In practice, II values in the range of 0.1–0.4 are typically considered highly informative and represent near-optimal performance, particularly in molecular systems where noise and imperfect correlations between descriptor spaces (e.g., structural features) and target observables (e.g., energy gaps or oscillator strengths) are unavoidable. Values between 0.4–0.7 still indicate a moderate degree of informativeness, while values in the range of 0.7–0.9 correspond to weak 
%but statistically meaningful 
correlations. Values approaching 1.0, by contrast, indicate an absence of meaningful information content.\cite{Donkor2023,Donkor2024,Wild2024,wild2025automatic,DiPino2025}

In the present work, the DII framework is applied in a sequence of dimensionality-reduction steps aimed at quantifying how strongly different internal structural degrees of freedom encode variations in the electronic energy gaps sampled along the TSH trajectories. To efficiently explore the space of possible feature combinations, and to avoid the combinatorial growth associated with testing all possible subsets, we employ the forward greedy selection strategy introduced in Ref.~\citenum{glielmo2022ranking}. This search procedure is combined with the DII implementation provided in the \texttt{DADApy} Python library.\cite{glielmo2022dadapy}

\subsection{Coulomb Matrix Descriptors}

We start by computing the Coulomb matrix\cite{rupp2012fast_coulomb_matrix} ($\text{CM}$) between all atoms for the trajectories in a given electronic state. The Coulomb matrix is defined as:

\begin{equation}
  \text{CM}_{ij} =
  \begin{cases}
    0.5Z_i^{2.4} & \text{for } i=j \\
    \frac{Z_i Z_j}{R_{ij}} & \text{for } i \neq j
  \end{cases}
\end{equation}

The diagonal elements can be seen as the interaction of an atom with itself and are essentially a polynomial fit of the atomic energies to the nuclear charge $Z_i$. The off-diagonal elements represent the Coulomb repulsion between nuclei $i$ and $j$.

\subsection{Chemical Systems}

There are 5 systems under study, and these trajectories were published in Refs.\citenum{Barbatti2006,Barbatti2007,gonza2024_jctc_dftb_naf,gonza2024_jctc_dftb_motors}: Methaniminium Cation, Furan, L-glutamine (L-gln), L-pyroglutamine-ammonium (L-pyro), and Molecular Motor: 9-(2,4,7-trimethyl-2,3-dihydro-1H-inden-1-ylidene)-9H-fluorene. Fig.\ref{fig:systems} shows the chemical structures of the 5 systems studied in this work.

All the systems studied in this work were simulated in prior work done by some of the authors using TSH at the time-dependent density functional tight binding (TD-DFTB) level of theory and  their non-radiative decay mechanisms were validated against higher levels of electronic structure theory, either time-dependent density functional theory (TD-DFT) or multireference configuration interaction with single and double excitations (MR-CISD) depending on the system. TD-DFT is a widely used electronic-structure approach for computing excited-state properties, including excitation energies, oscillator strengths, and excited-state gradients. TD-DFTB is a semi-empirical approximation to TD-DFT that retains the same general framework while substantially reducing the computational cost, making it well suited for generating large ensembles of NAD trajectories.

%%%%%%%%%%%%%%%%%%%%%%%%
\subsection{Protocol for Analysis}\label{sec:protocol}
%%%%%%%%%%%%%%%%%%%%%%%%

%\begin{figure}[H]
%  \centering
%  \includegraphics[width=1.0\linewidth]{figs/Fig_Protocol.pdf}
%  \caption{}
%  \label{fig:protocol}
%\end{figure}

The main steps of the protocol are summarized in Figure \ref{fig:protocol}. Below we outline important technical aspects of each step. The procedure begins with input data coming from NAD trajectories such as the positions of all particles as a function of time. In addition, in NAD simulations, one typically has the evolution of other physical quantities associated with the optical properties such as the electronic energy gap and the oscillator strength.

\begin{enumerate}

\item Extract the Coulomb Matrix ($\text{CM}_{ij}^{S_l}$) for each frame of an ensemble of NAD trajectories in a given electronic excited state, $S_l$, for each pair of atoms $i$ and $j$. Accumulate the results over all the frames to make a probability distribution of the elements of the Coulomb Matrix $P^{S_l}$.
%Assuming that the system hops from $S_l$ to $S_{l-1}$, where the latter may or may not be the ground state ($S_0$), 
We also compute the Coulomb Matrix for all the frames in the state the system hops to ($\text{CM}^{S_{l-1}}$), and the corresponding probability distribution of all pairwise Coulomb Matrix values, $P(\mathrm{CM}_{ij}|S_l)$ and $P(\mathrm{CM}_{ij}|S_{l-1})$. In general the system can hop between any two electronic states ($S_l \rightarrow S_{l-1}$). Specifically for the CI that we will focus on in this work, we look at transitions from $S_1 \rightarrow S_0$.

%GO AWAY DEB!!! COME BACK IN 2 HRS!! 

\item Compute the Jensen-Shannon divergence (JSD), which is a symmetrized version of the Kullback–Leibler Divergence (KLD)\cite{kullback1951information}, between $P(\mathrm{CM}_{ij}|S_l)$ and $P(\mathrm{CM}_{ij}|S_{l-1})$.

The KLD, which is a measure of similarity between 2 probability distributions $R$ and $Q$, is defined as
    $$
    D_\text{KL} (R||Q) = \sum_{x \in X} R(x) \log \frac{R(x)}{Q(x)}
    $$
Instead, the JSD is defined so as to symmetrize the KLD
    $$
    D_\text{JS} = 0.5 (D_\text{KL} (R || M) + D_\text{KL} (Q || M)), \text{where } M = 0.5 (R+Q)
    $$

\include{figs/flowchart_v2}

The outcome of determining the JSD is that it gives us ``hotspots'' of \emph{pairwise atomic interactions} (i.e., Coulomb Matrix elements) that change significantly from one state to another. This allows us to do a first dimensionality reduction step selecting only these ``hotspot'' atoms . We do this hotspot selection based on a threshold value of the JSD that we select such that the hotspots are those values where the $\text{JSD} > \text{threshold}$. These selected Coulomb Matrix hotspots are labeled as $\text{CM}_{\text{hotspot}}^{S_l}$. \text{threshold} is an adjustable parameter.

\item Next, we compute the DII($\text{CM}_{\text{hotspot}}^{S_l} \rightarrow \Delta E_{l, l-1})$, for all atoms $i, j$ belonging to the ``hotspots'', assuming we are analysing the NAD trajectory where the system jumps from $S_l$ to $S_{l-1}$. In addition, where the oscillator strength ($f^{osc}$) is available, we also compute DII($\text{CM}_{\text{hotspot}}^{S_l} \rightarrow f^{osc}_{l, l-1})$.

We filter the results of these DII calculations to identify a small set of Coulomb-matrix elements, and therefore the corresponding atom pairs, whose structural variations are most informative of changes in the electronic energy gaps. 
The filtering is done by selecting the $\texttt{nfeatures}$ (default value selected in this work is 10) values with the least DII. We denote these \textit{filtered} hotspots from the Coulomb Matrix as  $\text{CM}^{S_l}_\text{f}$. 

These coordinates describe nuclear motions that modulate the relative electronic-state energies during excited-state relaxation. When such motions lead to closure of the relevant energy gap, they provide a dynamical signature of approach toward a CI region. Oscillator strengths are then analyzed as a complementary electronic target, allowing us to determine which coordinates govern the intensity of the emission. 
%that is the approach to the conical intersection along the NAD trajectories, as well as the oscillator strength. 

\item  While the Coulomb-matrix analysis identifies the atom pairs that are most informative of the target electronic observable, these descriptors are not always the most chemically interpretable representation of the underlying motion. We therefore map the selected atoms onto an internal-coordinate representation, which allows the relevant structural changes to be expressed in terms of bonds, angles, and dihedrals.

Next, for all frames belonging to a given electronic state $S_l$, we generate the Z-matrix representation $\mathbf{R}^{S_l}$ using a combination of the quantum chemistry software ORCA\cite{orca_neese2012,orca_neese2018,orca_neese2020,orca_neese2022} and an in-house python script, which contains the corresponding internal coordinates, including bonds, angles, and dihedrals. To ensure that the motions involving the atoms selected in the Coulomb-matrix step are explicitly represented, we augment this set with additional internal coordinates constructed from those atoms when they are not already included in the default Z-matrix. 
%Next we generate the Z-matrix ($\textbf{R}^{S_l}$) for all frames belonging to a given electronic state $S_l$, which has information on all the internal coordinates (including bonds, angles, and dihedrals), and \emph{add to this other internal coordinates that may involve the selected atoms from the previous step}. 
Then we compute DII($ \textbf{R}_k^{S_l} \rightarrow \text{CM}_\text{f}^{S_l} $), i.e., we compute the DII for every internal coordinate ($\textbf{R}^{S_l}_k$) to the Coulomb Matrix hotspots that best explain the energy gaps/oscillator strengths. This allows us to select a subset of internal coordinates ($\textbf{R}^{S_l}_\text{selected}$). As in the previous step, here also we do the feature selection such that a specified $\texttt{nfeatures}$ (set to 10 by default) of the most informative coordinates are selected as the $\textbf{R}^{S_l}_\text{selected}$.

\item We subsequently do a final round of DII calculations using the selected internal coordinates to determine which ones best predict the energy gaps/oscillator strengths - DII($ \textbf{R}^{S_l}_\text{selected} \rightarrow \Delta E_{l, l-1})$ and DII($ \textbf{R}^{S_l}_\text{selected} \rightarrow f^{osc}_{l, l-1})$. This gives us a set of physically interpretable modes that are most informative on the energy gaps and oscillator strengths.
  
  %\item In addition, to compare to previous approaches and similar methods, we also compute the DII($ \text{PC}^{S_l}_n \rightarrow \Delta E_{l, l-1})$ and DII($ \text{PC}^{S_l}_n  \rightarrow f^{osc}_{l, l-1})$, between the principal components (PCs - where $\text{PC}^{S_l}_n$ represents the n-th PC from the trajectory in a given excited state) which gives us a series of ``collective" modes that can be used to explain the energy gaps and oscillator strengths.
\end{enumerate}

One motivation for adopting this multistep procedure, rather than directly computing the DII between all internal coordinates and the energy gaps, is to reduce the dimensionality of the candidate feature space. While in principle DII can directly infer the modes from a much larger and highly redundant set of internal coordinates, it needs a number of samples that grows exponentially with the number of features to perform feature selection in this space. This would be computationally intractable, particularly for larger systems. 
%In principle, one could define a much larger and highly redundant set of internal coordinates than those considered here; however, performing feature selection over such an expanded set would be computationally demanding and potentially prohibitive, particularly for larger systems. 
%One reason to go through these series of steps as opposed to computing the DII between internal coordinates and the energy gaps at the beginning is that this allows us to reduce the dimensionality of the space of the possible coordinates we consider since one can define a much larger list of redundant internal coordinates than what we consider here, and doing a feature selection over such a large list might be computationally prohibitive, especially for larger systems.

%%%%%%%%%%%%%%%%%%%%%%%%
\section{Results}\label{sec:results}
%%%%%%%%%%%%%%%%%%%%%%%%
In this section, we apply the workflow to a variety of systems of diverse photochemistry. We begin with two benchmark systems that have been extensively studied in photochemistry using NAD simulations, methaniminium cation and furan, to test whether the selected coordinates recover established non-radiative decay mechanisms. We then analyze two amino-acid systems,  
extensively studied in our group in the context of non-aromatic fluorescence (NAF)\cite{arnon2021off,kumar2022role,stephens2021short,pinotsi2016proton,Grisanti_JACS_2020,morzan_2022_jpcb,CO_lock_2023,Banerjee2025},
L-glutamine (L-gln) and L-pyroglutamine-ammonium (L-pyro), where we examine descriptors for both the energy gap and the oscillator strength, $f^{\mathrm{osc}}$. 
Finally, we consider an overcrowded alkene molecular motor, a light-driven molecular machine in which photoexcitation is converted into directional rotation around a central double-bond axle.\cite{Conyard2012,Kazaryan2011,Garcia2020} Because motor function depends on the non-radiative relaxation from the excited state to the ground state, this system provides a chemically motivated test case for identifying the coordinates that govern both non-radiative decay and bright-to-dark state conversion.\cite{Pang2017,Wen2023,gonza2024_jctc_dftb_motors}
%Finally, we study molecular motors to probe highly collective photochemistry and identify coordinates governing both non-radiative decay and bright-to-dark state conversion.

%%%%%%%%%%%%%%%%%%%%%%%%
\subsection{Methaniminium Cation}

%%%%%%%%%%%%%%%%%%%%%%%%
Methaniminium cation (\ce{CH2NH2+}) serves as a minimal model for protonated Schiff bases which are critical in the photophysics of retinal\cite{yang2022quantum} and has been investigated with several electronic-structure methods.\cite{Barbatti2006,Barbatti2007,Hollas2018,Suchan2020} We analyze TSH trajectories computed with TD-DFTB in a previous work published by our group.\cite{gonza2024_jctc_dftb_motors} 
%Consequently, the inferred non-radiative decay modes are expected to be closest to the TD-DFTB description, with subtle quantitative differences relative to MR-CISD-based trajectories.\cite{Barbatti2006, Barbatti2007}

\begin{figure}[H]
  \centering
  \includegraphics[width=1.0\linewidth]{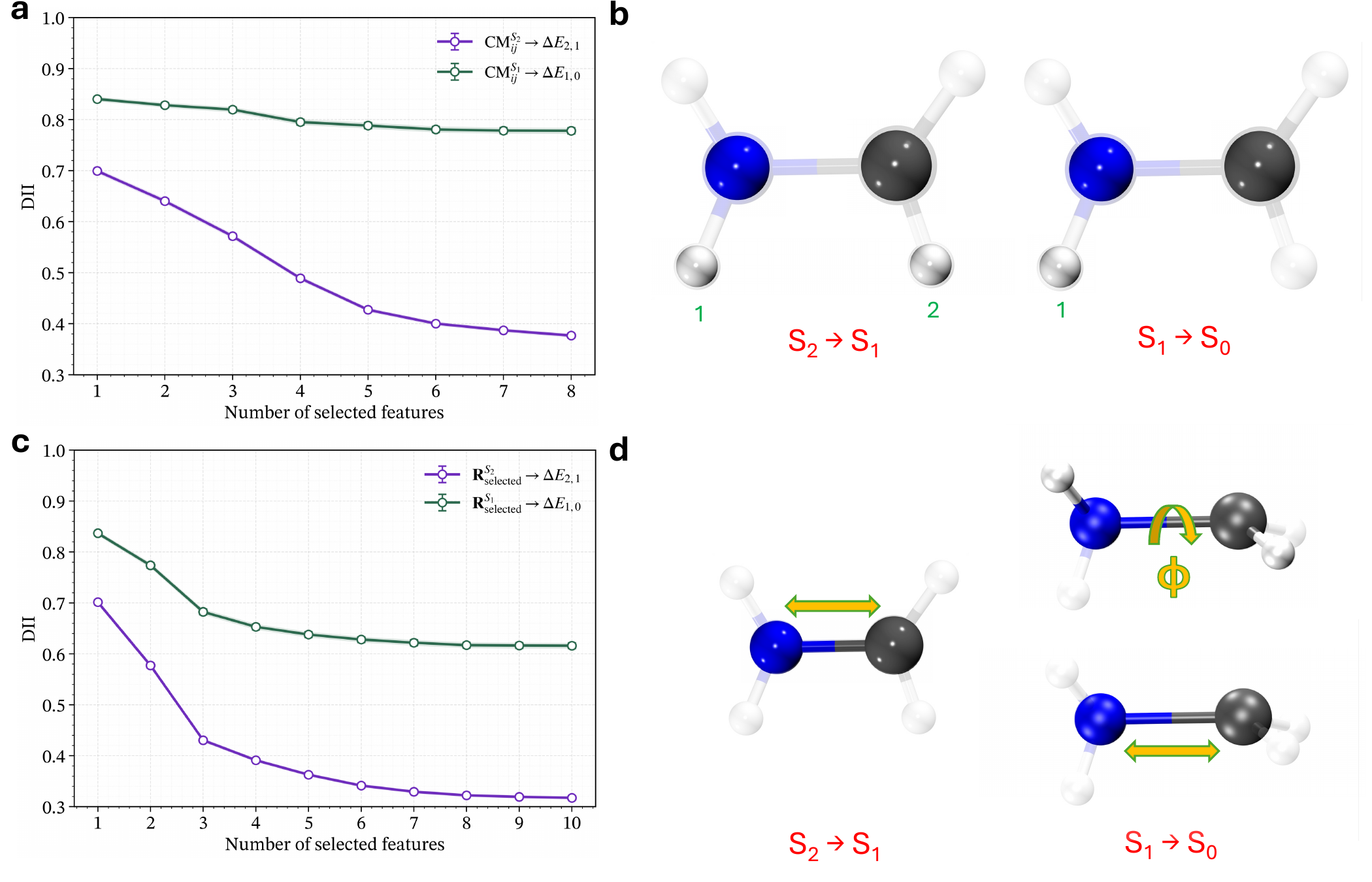}
\caption{
(a) DII between Coulomb-matrix descriptors and energy gaps as a function of the number of selected features. Purple corresponds to the S$_2 \rightarrow$ S$_1$ and green corresponds to the S$_1 \rightarrow$ S$_0$ relaxation channels respectively.
(b) Most informative atom pairs (“hotspots”) identified from Coulomb matrix analysis are highlighted. The left and right parts of the panel corresponds to the S$_2 \rightarrow$ S$_1$ and S$_1 \rightarrow$ S$_0$ channels respectively. 
(c) DII between internal coordinates and energy gaps as a function of the number of selected features. Purple corresponds to the S$_2 \rightarrow$ S$_1$ and green corresponds to the S$_1 \rightarrow$ S$_0$ relaxation channels respectively.
(d) The most informative structural modes extracted from the DII analysis: The S$_2 \rightarrow$ S$_1$ transition is governed primarily by C–N stretching (left), while the S$_1 \rightarrow$ S$_0$ channel involves the \ce{H(C)-C-N-H(N)} torsion and \ce{C-N} stretching (right).  
}
  \label{fig:formaldimine}
\end{figure}

%This is a three-state problem (ground state plus two excited states), 
The photochemistry of \ce{CH2NH2+} primarily involves studying the excitation to the $S_2$ state and the subsequent non-radiative decays to $S_1$ and then $S_0$. The photochemistry of this system is well known to be described by these three states.\cite{Barbatti2006,Barbatti2007}
We thus analyze two non-radiative channels: $S_2 \rightarrow S_1$ and $S_1 \rightarrow S_0$. In Fig.~\ref{fig:formaldimine}a we report DII$(\text{CM}_{ij}^{S_2} \rightarrow \Delta E_{2,1})$ (purple) and DII$(\text{CM}_{ij}^{S_1} \rightarrow \Delta E_{1,0})$ (green), i.e., how well selected Coulomb-matrix (CM) atom pairs explain each energy gap. The $S_2 \rightarrow S_1$ curve decreases substantially faster, indicating that a small number of pairwise descriptors already captures much of the relevant information for $\Delta E_{2,1}$, whereas $\Delta E_{1,0}$ requires a larger set of pairwise descriptors and the features that are selected are less informative on the energy gap. Figure~\ref{fig:formaldimine}b identifies the dominant pairs: for $S_2 \rightarrow S_1$, \{C,N\}, \{N,H2\}, and \{C,H1\}; for $S_1 \rightarrow S_0$, \{C,N\} remains important, with \{N,H1\} and \{C,H1\} completing the top three. These are the aforementioned ``hotspots'' identified in Step 2 of the analysis protocol.

We then translate these pair-level signatures into chemically interpretable internal coordinates according to Steps 4 and 5 in the analysis protocol. In Fig.~\ref{fig:formaldimine}c, only 2--3 features are sufficient to approach the minimum DII for $S_2 \rightarrow S_1$, while the $S_1 \rightarrow S_0$ curve decreases more slowly and saturates at a higher DII. The top coordinates are shown in Fig.~\ref{fig:formaldimine}d. For $S_2 \rightarrow S_1$, the leading mode is clearly \ce{C-N} stretching, followed by a torsion around the \ce{C-N} bond (SI Fig.~\ref{fig_SI:formal_extra_modes}a). This agrees with earlier work identifying the \ce{C-N} stretch as the dominant coordinate for the $S_2/S_1$ crossing. In contrast, for $S_1 \rightarrow S_0$, the most informative coordinate is the \ce{H(C)-C-N-H(N)} dihedral, with \ce{C-N} stretching as a secondary contributor and \ce{N-CH2} pyramidalization appearing as the next relevant degree of freedom (SI Fig.~\ref{fig_SI:formal_extra_modes}b).

The single-feature DII values in Fig.~\ref{fig:formaldimine}c (about 0.70 for $S_2/S_1$ and about 0.85 for $S_1/S_0$) further support this asymmetry: one dominant coordinate explains the upper-state decay significantly better than it explains the lower-state decay. As additional coordinates are included, the stronger drop for $S_2 \rightarrow S_1$ and the more modest improvement for $S_1 \rightarrow S_0$ indicate that the latter process is intrinsically more collective. This is consistent with previous mechanistic analyses reporting that access to the $S_1/S_0$ seam involves coupled torsion, \ce{C-N} stretching, and pyramidalization-type distortions of the \ce{CH2} group.\cite{Barbatti2006,Barbatti2007}. In summary, these findings demonstrate that our protocol provides a semi-automated manner for diagnosing relevant photochemical decay pathways in a well studied system involving non-adiabatic hops associated with three electronic states.

%%%%%%%%%%%%%%%%%%%%%%%%
\subsection{Furan}
\begin{figure}[H]
  \centering
  \includegraphics[width=1.0\linewidth]{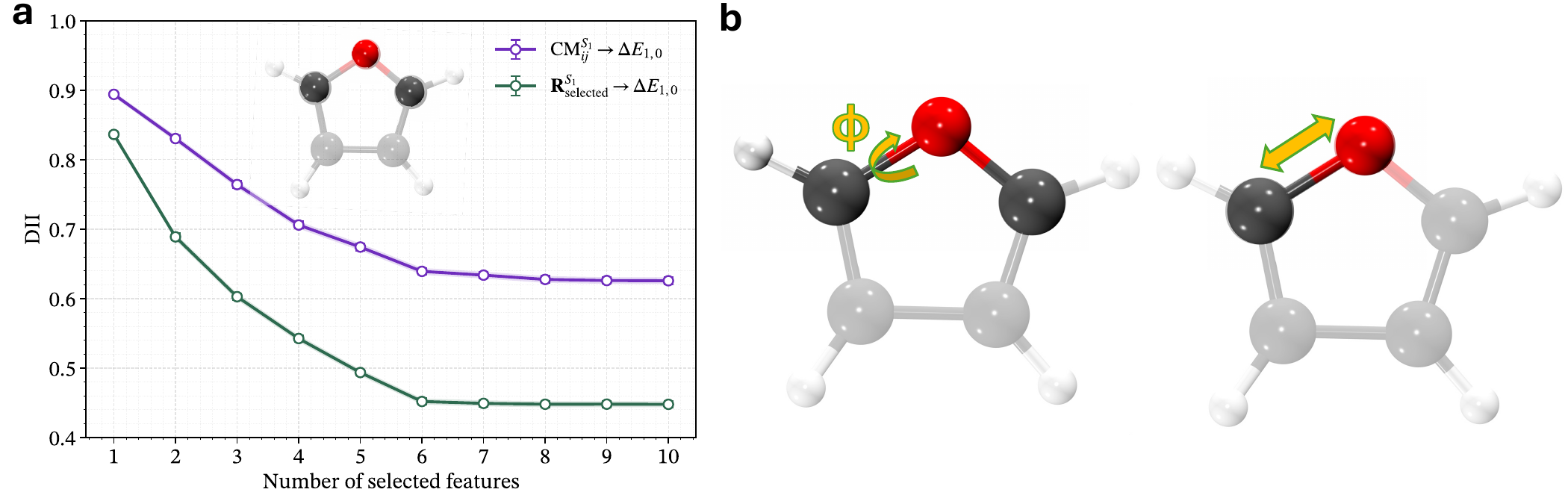}
  \caption{
(a) DII between Coulomb-matrix descriptors and the S$_1 \rightarrow$ S$_0$ energy gap as a function of the number of selected features (purple) and DII between internal coordinates and S$_1 \rightarrow$ S$_0$ energy gap as a function of the number of selected features (green). The inset shows the most informative atom pairs (“hotspots”) identified from Coulomb matrix analysis are highlighted: the O atom and the two adjacent C atoms.
(b) The most informative structural modes extracted from the DII analysis: the \ce{C-O-C-H} dihedral and a \ce{C-O} stretching coordinate.  
  }
  \label{fig:furan}
\end{figure}

%%%%%%%%%%%%%%%%%%%%%%%%
Furan is another canonical benchmark for non-adiabatic dynamics and was also investigated in a previous study from our group.\cite{gonza2024_jctc_dftb_motors} Our analysis focuses on the $S_1/S_0$ non-radiative decay channel. Figure~\ref{fig:furan}a (purple curve) reports DII$(\text{CM}_{ij}^{S_1} \rightarrow \Delta E_{1,0})$ as a function of the number of selected Coulomb-matrix (CM) features. This decreases gradually, from about 0.89 to about 0.63 over the first eight features, indicating that no single atom pair dominates the gap fluctuations. The most informative CM pairs are shown as the inset in Fig.~\ref{fig:furan}a: one \{C,C\} and two \{C,O\} pairs, emphasizing that the relevant structural signal is distributed over the heteroatom-containing ring framework.
%In higher-level descriptions, two low-lying excited states are typically discussed for this system: a Rydberg ($\pi3s$) state and a valence $\pi\pi^*$ state.\cite{Fuji2010,Oesterling2017} Since TD-DFTB with a minimal basis does not describe the Rydberg state, our analysis focuses on the $S_1/S_0$ channel where $S_1$ has predominantly $\pi\pi^*$ character.

Moving to the chemically interpretable internal coordinates (green curve in Fig.~\ref{fig:furan}a), we observe that the ranked internal-coordinate representation lowers DII from about 0.84 to about 0.45 only after including six modes, again pointing to a collective decay coordinate. The leading interpretable modes are shown in Fig.~\ref{fig:furan}b: a \ce{C-O-C-H} dihedral and a \ce{C-O} stretching coordinate. SI Fig.~\ref{fig_SI:furan_extra_modes} shows the next-ranked contributions, including the second \ce{C-O} stretch, the \ce{C-O-C} angle, and a \ce{C-C} distance; an explicit ring-puckering coordinate also appears among the top six selected features. Altogether, these modes describe coupled out-of-plane deformation (the \ce{C-O-C-H} dihedral and  \ce{C-O-C} angle are most notably correlated to the ring puckering) and ring-opening (the \ce{C-O} and \ce{C-C} stretches), consistent with prior theoretical studies that associate furan's non-radiative decay with puckering and ring-opening pathways in a multidimensional photochemical landscape.\cite{Fuji2010,Oesterling2017,gonza2024_jctc_dftb_motors}.

%%%%%%%%%%%%%%%%%%%%%%%%
\subsection{L-glutamine}
%%%%%%%%%%%%%%%%%%%%%%%%

The preceding two systems demonstrate that our protocol can identify non-radiative decay modes in well-studied prototypical molecules with different numbers of electronic states and active nuclear degrees of freedom. 
%We next turn to amino-acid-based systems associated with NAF, focusing first on L-glutamine (L-gln). Experimentally, L-gln is non-fluorescent and decays non-radiatively to the ground state in a short time.
We next turn to amino-acid-based systems associated with NAF, beginning with L-glutamine (L-gln). Experimentally, L-gln is non-fluorescent, consistent with efficient non-radiative relaxation back to the ground state on a short timescale.
%; however, upon heating in water and complexation with an ammonium ion it undergoes a chemical transformation that yields L-pyroglutamine-ammonium (L-pyro-amm), a distinct supramolecular structure stabilized by short hydrogen bonds and characterized by enhanced fluorescence.\cite{stephens2021short,CO_lock_2023} 
Motivated by this connection between structural constraints, and different non-radiative decay mechanisms and fluorescence properties, we analyze not only the energy gap $\Delta E_{1,0}$ but also the oscillator strength $f^{\mathrm{osc}}_{1,0}$ along the trajectories. This allows us to determine if the oscillator strength can be explained accurately by simple structural coordinates.

\begin{figure}[H]
  \centering
  \includegraphics[width=1.0\linewidth]{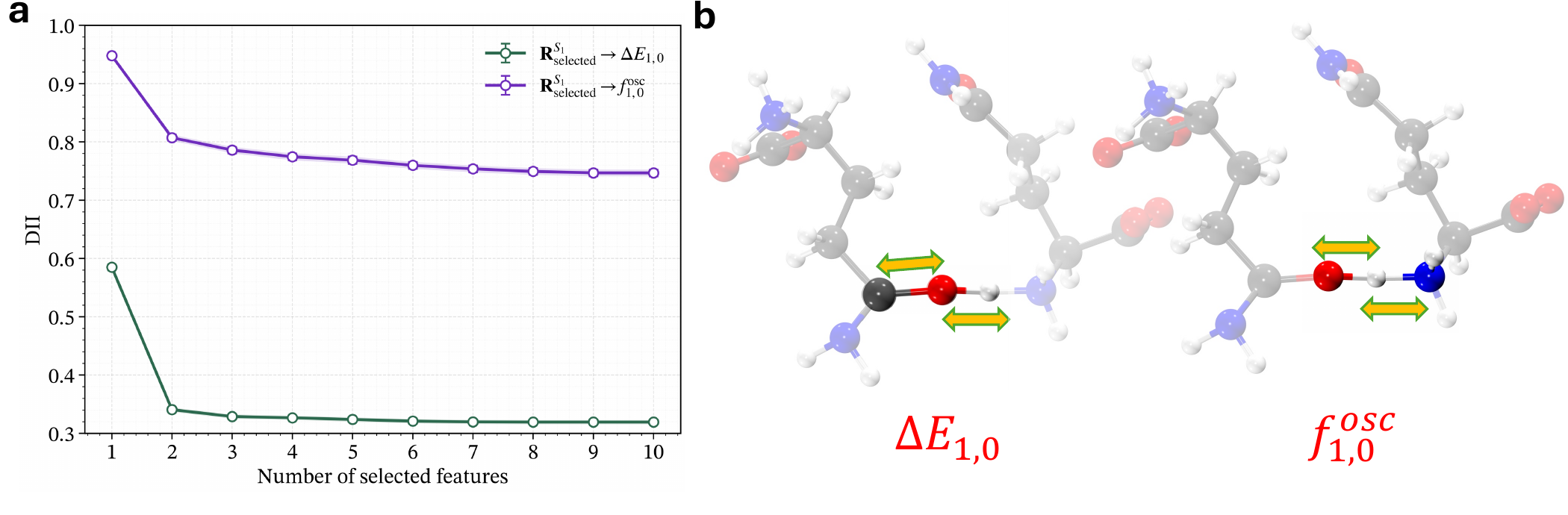}
  \caption{
(a) DII between internal coordinates and energy gap/oscillator strength as a function of the number of selected features. Purple corresponds to the $f^{\mathrm{osc}}_{1,0}$ and green corresponds to the $\Delta E_{1,0}$ respectively.
(b) The most informative structural modes extracted from the DII analysis: The $\Delta E_{1,0}$ is governed primarily by the carbonyl \ce{C-O} stretch and an \ce{O-H} distance (left), while the $f^{\mathrm{osc}}_{1,0}$ is described by the \ce{O-H} and \ce{H-N} stretching (right).  
  }
  \label{fig:gln}
\end{figure}

In SI Fig.~\ref{fig_SI:gln_cmat}a we compare DII trends for two targets, $\Delta E_{1,0}$ (green) and $f^{\mathrm{osc}}_{1,0}$ (purple), using selected Coulomb-matrix (CM) elements. The contrast is striking: for the energy gap, DII drops from about 0.58 to about 0.32 with only two selected features and then plateaus, whereas for oscillator strength it remains much higher (about 0.95 to about 0.72 over 10 features). SI Fig.~\ref{fig_SI:gln_cmat}b shows the corresponding key atom pairs: for $\Delta E_{1,0}$, the dominant signal involves the carbonyl atoms plus one neighboring \ce{NH3+} hydrogen; for $f^{\mathrm{osc}}_{1,0}$, the carbonyl oxygen and all three \ce{NH3+} hydrogens are most implicated.

The same hierarchy appears with chemically interpretable internal coordinates. In Fig.~\ref{fig:gln}a we display the DII results for both $\Delta E_{1,0}$ (green) and $f^{\mathrm{osc}}_{1,0}$ (purple) using the internal coordinates. Fig.~\ref{fig:gln}b shows that the top two descriptors are the carbonyl \ce{C-O} stretch and an \ce{O-H} distance associated with proton transfer; together they already recover the low-DII regime (\(\sim 0.32\)). In fact, the single feature associated with the \ce{C-O} stretch already yields a DII of \(\sim 0.58\), indicating that this coordinate alone captures a substantial fraction of the structural information relevant to the energy gap. For the oscillator strength, the best descriptors are proton-transfer-related distances (\ce{O-H} and \ce{H-N}), but the DII stays near \(\sim 0.76\), indicating weaker direct structural predictability than for the energy gap. This is consistent with previous L-gln analyses, where non-radiative decay is dominated by localized carbonyl/proton-transfer motion.\cite{CO_lock_2023,gonza2024_jctc_dftb_naf}

The analysis of the oscillator strength is particularly noteworthy, as it highlights a distinct behavior compared to the energy gap. Specifically, the DII values remain significantly higher, indicating that the oscillator strength is governed by more complex relationships than those captured by the simple structural modes considered here. Nevertheless, moderate correlations are still observed. Importantly, the modes that are most informative for describing the energy gap are not necessarily identical to those relevant for $f^{\mathrm{osc}}$. 
%For example, the \ce{O-H} stretching mode plays a key role in predicting both $\Delta E_{1,0}$ and $f^{\mathrm{osc}}_{1,0}$, whereas the \ce{N-H} stretch is primarily relevant for the latter, and the \ce{C-O} stretch is more strongly associated with the former.

%%%%%%%%%%%%%%%%%%%%%%%%
\subsection{L-pyroglutamine-ammonium}
%%%%%%%%%%%%%%%%%%%%%%%%
%In contrast to L-gln, the L-pyroglutamine-ammonium complex (L-pyro) is non-aromatically fluorescent, 
In contrast to L-gln, L-pyroglutamine-ammonium (L-pyro) corresponds to the fluorescent product formed after chemical transformation of L-gln upon heating in water, as shown experimentally in previous work.\cite{stephens2021short} Subsequent nonadiabatic dynamics studies further showed that the short-hydrogen bonds  in L-pyro suppress access to the dominant non-radiative decay pathway by restricting the carbonyl-elongation mode and associated proton-transfer motion.\cite{CO_lock_2023,gonza2024_jctc_dftb_naf} As a result, only about \(20\%\) of trajectories in the present ensemble undergo non-radiative relaxation to the ground state.
%so non-radiative decay events are comparatively sparse in the trajectory ensemble. Here, only \(\sim 20\%\) of trajectories undergo hopping to lower states. 
This makes L-pyro a challenging case of a system with a limited amount of data because the analysis must recover meaningful decay coordinates from a limited number of reactive events.

\begin{figure}[H]
  \centering
  \includegraphics[width=1.0\linewidth]{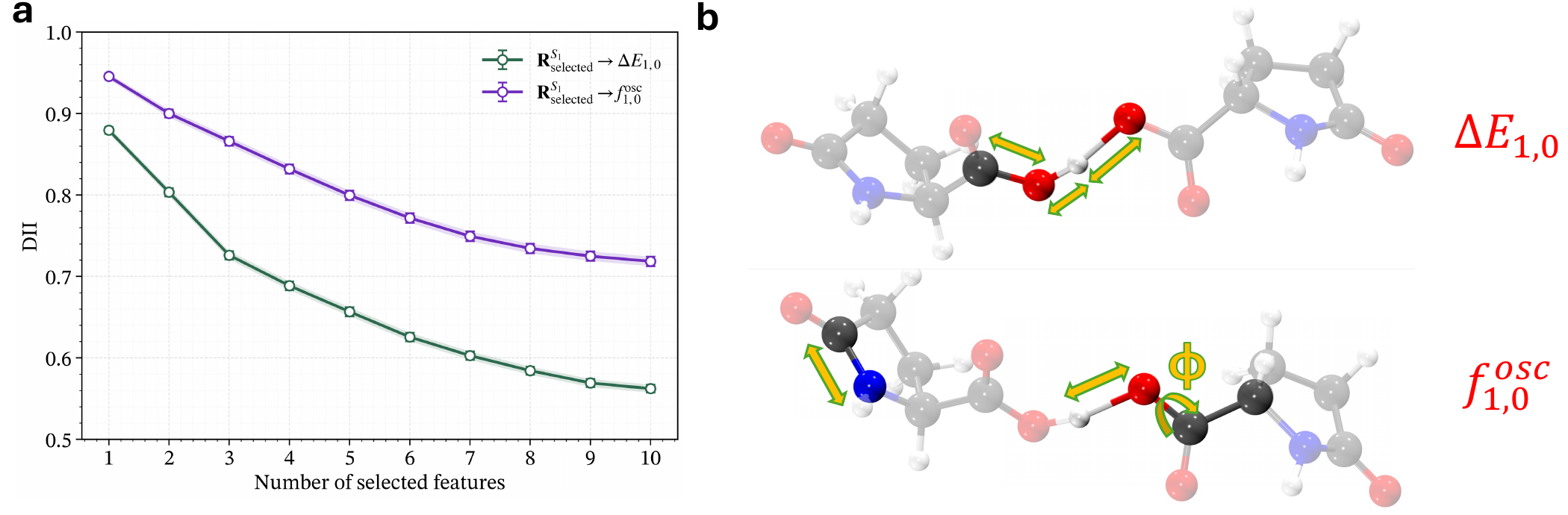}
  \caption{
(a) DII between internal coordinates and energy gap/oscillator strength as a function of the number of selected features. Purple corresponds to the $f^{\mathrm{osc}}_{1,0}$ and green corresponds to the $\Delta E_{1,0}$ respectively.
(b) The most informative structural modes extracted from the DII analysis: The $\Delta E_{1,0}$ is governed primarily by the carbonyl \ce{C-O} stretch and the proton transfer involving the two \ce{O-H} distances (top), while the $f^{\mathrm{osc}}_{1,0}$ is described by a \ce{O-H} stretch, a \ce{C-C-O-H} dihedral, and a \ce{C-N} ring stretch (bottom).  
  }
  \label{fig:pyro}
\end{figure}

In SI Fig.~\ref{fig_SI:pyro_cmat}a we analyze DII trends for both $\Delta E_{1,0}$ and $f^{\mathrm{osc}}_{1,0}$ using selected Coulomb-matrix (CM) features. Both curves decrease gradually, indicating pronounced collectivity. The energy gap remains more predictable than oscillator strength (DII approaches \(\sim 0.50\) for $\Delta E_{1,0}$ versus \(\sim 0.66\) for $f^{\mathrm{osc}}_{1,0}$), again showing that pairwise structural descriptors are less effective for optical observables. SI Fig.~\ref{fig_SI:pyro_cmat}b highlights atoms from the top five CM pairs; these are concentrated around carbonyl \ce{C-O} groups, the inter-dimer hydrogen-bond region, and the carbon atom adjacent to the carbonyl. Similar modes were identified by work done in our group in Ref.\citenum{CO_lock_2023}.

The internal-coordinate analysis in Fig.~\ref{fig:pyro}a gives a consistent picture. For $\Delta E_{1,0}$, the leading coordinates are the carbonyl \ce{C-O} stretch and the proton-transfer coordinate involving the two \ce{O-H} bonds (top panel of Fig.~\ref{fig:pyro}b). For $f^{\mathrm{osc}}_{1,0}$, the most relevant features shift toward proton-transfer and nearby torsional motions: an \ce{O-H} coordinate, a \ce{C-C-O-H} dihedral, and again a \ce{C-N} ring stretch (bottom panel of Fig.~\ref{fig:pyro}b). The slow DII saturation that we see for both $\Delta E_{1,0}$ and $f^{\mathrm{osc}}_{1,0}$) in Fig.~\ref{fig:pyro}a confirms that both observables are governed by coupled distortions rather than by a single dominant mode. 

As in the case of L-gln, the modes governing $f^{\mathrm{osc}}$ only partially overlap with those correlated with the energy gap. In particular, the \ce{O-H} stretching mode associated with proton transfer is common to both. In contrast, the \ce{C-N} ring stretch and the \ce{C-C-O-H} dihedral are uniquely relevant for $f^{\mathrm{osc}}$ and do not appear among the dominant modes describing the energy gap. This further reinforces the non-trivial observation that modes strongly correlated with the oscillator strength are not necessarily those that most effectively explain the energy gap.

We note that the oscillator strength ($f^{\mathrm{osc}}_{1,0}$) depends on both the energy gap ($\Delta E_{1,0}$) and the transition dipole moment (TDM). Given the partial overlap between the coordinates selected for the energy gap and those for the oscillator strength, it is natural to question whether this overlap arises primarily from the explicit dependence of $f^{\mathrm{osc}}$ on $\Delta E$. To assess this, we examined $\mathrm{DII}(\mathbf{R}^{S_1}_{\mathrm{selected}} \rightarrow f^{\mathrm{osc}}_{1,0}\mathrm{(TDM)})$, where $f^{\mathrm{osc}}_{1,0}\mathrm{(TDM)}$ denotes the oscillator strength constructed by retaining only the TDM contribution, excluding the energy gap term. For both L-gln and L-pyro, the resulting DII values are nearly identical to those obtained using the full oscillator strength (SI Fig.~\ref{fig_SI:gln_fosc_tdm} and SI Fig.~\ref{fig_SI:pyro_fosc_tdm}). In particular, the absolute magnitudes differ by less than $0.02$--$0.03$, the trends across the DII curves are also consistent, and the dominant physical modes remain unchanged relative to those identified from $\mathrm{DII}(\mathbf{R}^{S_1}_{\mathrm{selected}} \rightarrow f^{\mathrm{osc}}_{1,0})$. We will see in the next section however that this is not necessarily true in general.

%%%%%%%%%%%%%%%%%%%%%%%%
\subsection{Molecular Motors}
%%%%%%%%%%%%%%%%%%%%%%%%
Next, we analyze the overcrowded molecular motor\cite{Feringa2004},
%9-(2-methyl-2,3-dihydro-1H-cyclopenta[a]naphthalen-1-ylidene)-9H-fluorene
a light-activated system where photoexcitation initiates rotation around a central double-bond region.\cite{Garcia2020} 
%The trajectories were generated with TD-DFTB, as in our earlier work,\cite{gonza2024_jctc_dftb_motors} which enables statistically meaningful sampling for a system size that is challenging for more expensive electronic-structure methods.

This motor is also a useful test case because the key early-time photophysics combines two coupled observables: non-radiative relaxation and a bright-to-dark transition in $S_1$ (drop in oscillator strength) before ground-state recovery.\cite{Conyard2012} Previous studies implicate central-bond torsion, bond-stretching and pyramidalization in this process,\cite{Pang2017,Wen2023,gonza2024_jctc_dftb_motors} but disagree on which motion is the main driver.\cite{Conyard2012,Wen2023,Kazaryan2011} Our goal here is to identify the most informative (``parent'') coordinate directly from trajectory data.

\begin{figure}[H]
  \centering
  \includegraphics[width=1.0\linewidth]{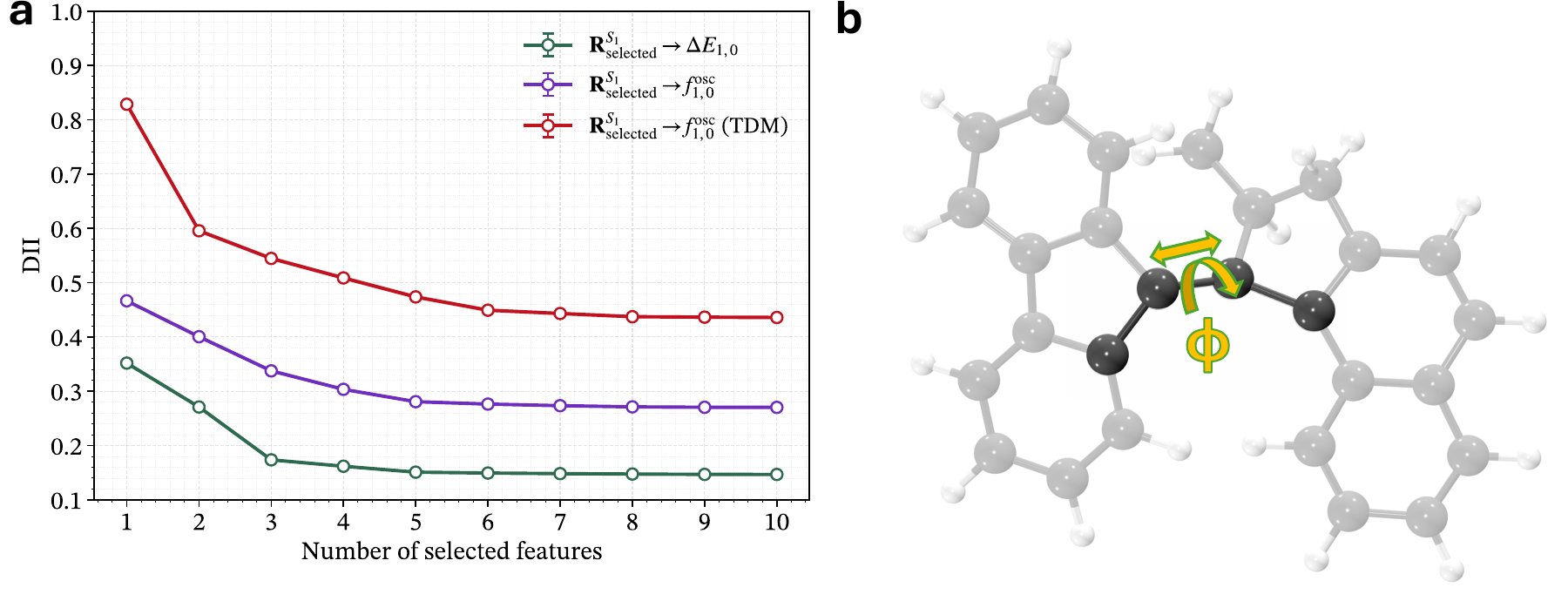}
  \caption{
(a) DII between internal coordinates and energy gap/oscillator strength as a function of the number of selected features. Purple corresponds to the $f^{\mathrm{osc}}_{1,0}$, red corresponds to $f^{\mathrm{osc}}_{1,0}\mathrm{(TDM)}$, and green corresponds to the $\Delta E_{1,0}$ respectively.
(b) The most informative structural modes extracted from the DII analysis: both the $\Delta E_{1,0}$ and the $f^{\mathrm{osc}}_{1,0}$ is governed primarily by the central torsion dihedral and the central double bond stretch that are highlighted.  
  }
  \label{fig:motors}
\end{figure}

In SI Fig.~\ref{fig_SI:motors_cmat}a, DII computed from selected Coulomb-matrix (CM) features decreases rapidly for both targets, reaching about \(\sim 0.27\) for $\Delta E_{1,0}$ and \(\sim 0.35\) for $f^{\mathrm{osc}}_{1,0}$. The most informative CM pairs are highlighted in SI Fig.~\ref{fig_SI:motors_cmat}b: for the energy gap, \{C1,C3\}, \{C1,C5\}, and \{C2,C6\}; for oscillator strength, \{C2,C4\}, \{C1,C3\}, and \{C1,C5\}. These pairs mainly track atoms whose relative geometry changes strongly along the central-bond rotation pathway.

We subsequently map this signal to interpretable internal coordinates. As shown in Figure~\ref{fig:motors}a with only three coordinates, DII already reaches very low values (about 0.17 for $\Delta E_{1,0}$ (green) and about 0.33 for $f^{\mathrm{osc}}_{1,0}$ (purple)), indicating that a compact coordinate set captures most of the relevant dynamics. The leading coordinates are the central torsional dihedral and the associated double-bond stretch (Fig.~\ref{fig:motors}b), with an additional smaller contribution from a second nearby dihedral (SI Fig.~\ref{fig_SI:motors_extra_modes}).

Quantitatively, the primary dihedral alone yields DII values of approximately 0.35 for $\Delta E_{1,0}$ and 0.46 for $f^{\mathrm{osc}}_{1,0}$, making it the single most informative coordinate for both non-radiative decay and bright-to-dark conversion. As in previous cases, however, the correlation between structural coordinates and the oscillator strength remains weaker, as reflected by the higher DII value. While the central dihedral captures the behavior of the energy gap nearly optimally, its correlation with the oscillator strength is comparatively less precise. The next most relevant contributions arise from the central double-bond stretch and an additional dihedral around the same region (SI Fig.~\ref{fig_SI:motors_extra_modes}), although their influence is clearly less significant.
In contrast to the L-gln and L-pyro systems, the modes governing the energy gap and oscillator strength in this system coincide. In both cases, the central torsional dihedral and the associated double-bond stretch emerge as the most informative coordinates, representing a notable scenario in which structural modes exhibit a relatively strong correlation with the oscillator strength. Taken together, these results support a mechanism in which torsional motion about the central bond acts as the primary driving coordinate, while the double-bond stretch and pyramidalization arise predominantly as secondary, coupled responses rather than primary triggers.

This system also highlights a practical strength of the protocol. The CM-filtering step alone does not explicitly return all atoms of the final parent dihedral, yet the subsequent internal-coordinate ranking resolves the dominant mechanistic mode unambiguously. In other words, the multi-step DII procedure can disentangle coupled photochemical coordinates even when the first feature-selection stage is not directly interpretable in terms of a single geometrical mode.

The strong overlap between the modes governing the energy gap and the oscillator strength that we observed in the previous section motivated a more detailed analysis of the contribution arising solely from the transition dipole moment (TDM), i.e., $f^{\mathrm{osc}}_{1,0}\mathrm{(TDM)}$. As before, we compute $\mathrm{DII}(\mathbf{R}^{S_1}_{\mathrm{selected}} \rightarrow f^{\mathrm{osc}}_{1,0}\mathrm{(TDM)})$, shown as the red curve in Fig.~\ref{fig:motors}a.
In contrast to the behavior observed for L-gln and L-pyro, the DII in this case saturates at a significantly higher value ($\sim 0.44$) compared to that obtained using the full oscillator strength (purple curve), which reaches $\sim 0.27$. Similarly, the top three coordinates yield a DII of only $\sim 0.54$ for $f^{\mathrm{osc}}_{1,0}\mathrm{(TDM)}$, as opposed to $\sim 0.33$ for the full $f^{\mathrm{osc}}_{1,0}$. Notably, these top three coordinates are identical to those identified previously, and appear in the same order; however, their ability to describe the TDM-only contribution is markedly reduced relative to their performance for the full oscillator strength.
This behavior contrasts quite significantly with the L-gln and L-pyro systems, where $f^{\mathrm{osc}}_{1,0}\mathrm{(TDM)}$ closely mirrors the behavior of $f^{\mathrm{osc}}_{1,0}$. The present results therefore indicate that, in this molecular motor system, the energy gap plays a more dominant role in determining the oscillator strength than in the amino acid systems considered earlier. More generally, repeating our protocol with other types of electronic observables coming from population analysis in quantum chemistry\cite{orca_neese2012,orca_neese2018,orca_neese2020,orca_neese2022} where the electron density is partitioned into atomic contributions, could be interesting to explore in the future within this context. 

%%%%%%%%%%%%%%%%%%%%%%%%
\subsection{Non-Radiative Decay Pathways Along Principal Components}
%%%%%%%%%%%%%%%%%%%%%%%%
%\begin{figure}[H]
%  \centering
%  \includegraphics[width=1.0\linewidth]{figs/Fig_PCA.pdf}
%  \caption{Principal component analysis of the L-gln (a) and L-pyro (b) systems, showing the DII between the leading PCs and the corresponding observables.}
%  \label{fig:pca_comparison}
%\end{figure}

\begin{figure}[H]
  \centering
  \includegraphics[width=1.0\linewidth]{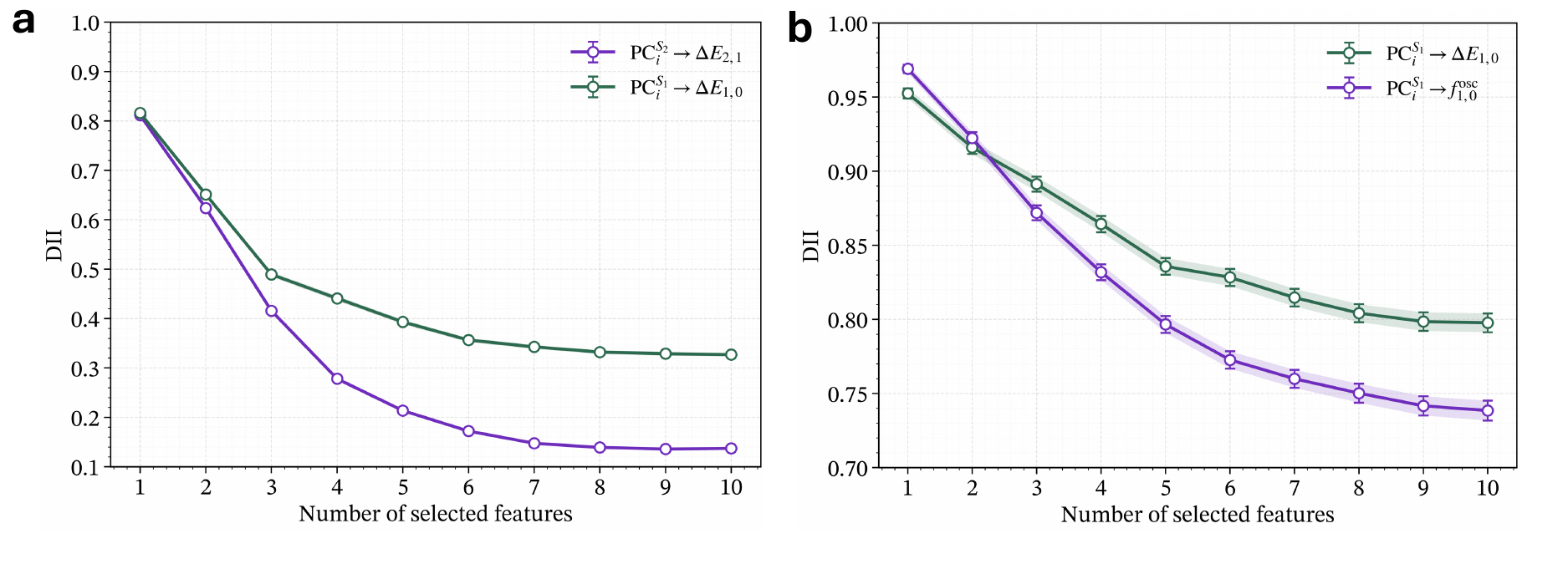}
  \caption{
(a) Methaniniminium Cation: DII between the leading PCs and energy gaps as a function of the number of selected features. Purple corresponds to the S$_2 \rightarrow$ S$_1$ and green corresponds to the S$_1 \rightarrow$ S$_0$ relaxation channels respectively.
(b) L-gln: DII between the leading PCs and energy gap/oscillator strength as a function of the number of selected features. Purple corresponds to the $f^{\mathrm{osc}}_{1,0}$ and green corresponds to the $\Delta E_{1,0}$ respectively.
  }
  \label{fig:pca_comparison}
\end{figure}

The internal-coordinate analysis presented above starts from a broad structural representation and systematically selects the coordinates that are most informative of the target electronic observables. In some of the systems we studied (L-gln and Molecular Motors), this procedure leads to a rapid decrease of the DII, indicating that a small number of localized coordinates is sufficient to capture most of the relevant structural information. However, in other systems (Furan and L-pyro), 
%particularly for oscillator strengths and for the larger molecular systems considered here, 
the DII decreases more gradually as additional coordinates are included. This behavior suggests that the changes in the energy gap may depend on a more distributed nuclear distortion rather than on a small set of localized bonds, angles, or dihedrals.

To assess this possibility, we complement the internal-coordinate analysis with a principal component (PC) based representation of the trajectory ensemble. Principal components provide a linear, variance-maximizing basis for nuclear motion and therefore offer a compact description of correlated, delocalized fluctuations in configuration space.\cite{Buchner1992} 
We therefore compute the DII between PC coordinates and the relevant electronic observables across all systems, using PCA not as an alternative to the chemically interpretable coordinate selection protocol, but as a reference representation for evaluating the extent to which the observables are controlled by collective nuclear motion.
%To assess the role of collective nuclear motions, we computed the DII between principal component (PC) coordinates and the relevant electronic observables across all systems. Principal components provide a linear, variance-maximizing representation of the nuclear motion, and are therefore expected to capture correlated, delocalized fluctuations in trajectory space.\cite{Buchner1992} This makes them a natural reference for comparison with the physically motivated internal-coordinate representations used throughout this work. 
Compared to the final step of the protocol where we use internal coordinates, when using PCs we instead compute DII$(\mathrm{PC}_{i}^{S_l} \rightarrow \Delta E_{l,l-1})$ and DII$(\mathrm{PC}_{i}^{S_l} \rightarrow f^{\mathrm{osc}}_{l,l-1})$ for each $\mathrm{PC}_{i}$, to determine the PCs that contribute most to the energy gap and the oscillator strength respectively. 

For the methaniminium cation, projecting the dynamics onto PCs substantially enhances predictability. In particular, DII$(\mathrm{PC}_{i}^{S_2} \rightarrow \Delta E_{2,1})$ and DII$(\mathrm{PC}_{i}^{S_1} \rightarrow \Delta E_{1,0})$ decrease rapidly with the number of retained components 
%(SI Fig.~\ref{fig_SI:formal_pc_dii})
(Fig.~\ref{fig:pca_comparison}a)
, with the first $\sim$10 PCs nearly saturating the predictability of the $S_2/S_1$ gap. A similar trend is observed for $\Delta E_{1,0}$, where the DII is reduced to $\sim 0.32$, significantly below the $\sim 0.60$ obtained using the top internal coordinates. This demonstrates that collective combinations of motions encode a large fraction of the information relevant to nonadiabatic relaxation, consistent with previous observations that dimensionality reduction techniques can recover dominant reaction pathways directly from trajectory data without prior mechanistic assumptions.\cite{Li2017,Zhu2024}  Similar trends are observed in Furan (see caption of SI Fig.~\ref{fig_SI:furan_pc_dii} for a more detailed summary).

%A comparable behavior is observed for furan (SI Fig.~\ref{fig_SI:furan_pc_dii}), where the inclusion of PCs continues to lower the DII to $\sim 0.32$. This indicates that, even when a select number of chemically intuitive coordinates already captures part of the dynamics, additional predictive information is contained in more delocalized collective modes. Such behavior reflects the inherently high-dimensional and strongly coupled nature of nonadiabatic dynamics, where multiple nuclear degrees of freedom contribute simultaneously to the evolution toward conical intersections.

In contrast, the amino-acid systems reveal a qualitatively different regime. As shown in Fig.~\ref{fig:pca_comparison}b, for L-gln the PC representation performs significantly worse for the energy gap $\Delta E_{1,0}$, with the DII saturating near $\sim 0.80$, compared to $\sim 0.32$ obtained using only two interpretable internal coordinates. This clearly indicates that the dominant relaxation coordinate is highly localized, and that its signal is diluted when embedded into delocalized PC directions. For the oscillator strength $f^{\mathrm{osc}}_{1,0}$, PCs provide only a marginal improvement ($\sim 0.73$ versus $\sim 0.76$), suggesting that this observable has a somewhat more collective character but still does not benefit strongly from a purely variance-based representation. This breakdown of PCA highlights a known limitation: while PCs optimally capture structural variance, they do not necessarily align with the physically relevant directions governing electronic transitions. In L-pyro 
%(Fig.~\ref{fig:pca_comparison}b)
(SI Fig.~\ref{fig_SI:pyro_pc_dii})
, PCs noticeably improve the mapping to the oscillator strength (DII decreases from $\sim 0.72$ to $\sim 0.63$), indicating that $f^{\mathrm{osc}}_{1,0}$ is influenced by more distributed structural rearrangements. However, the improvement for the energy gap is minimal ($\sim 0.56$ to $\sim 0.54$).

Finally, for the molecular motors (SI Fig.~\ref{fig_SI:motors_pc_dii}), PC-based descriptors achieve low DII values ($\sim 0.29$ for the energy gap and $\sim 0.33$ for oscillator strength), confirming that collective motions are highly informative in these larger, more flexible systems. These values however, remain slightly higher than those obtained using targeted internal coordinates, again pointing to the presence of specific mechanistic directions that are not optimally aligned with variance-maximizing modes.

Taken together, these results establish a clear dichotomy. Principal components are highly effective at compressing high-dimensional trajectory data and can capture substantial predictive information, particularly in systems where the dynamics is intrinsically collective. This is consistent with their widespread use in analyzing nonadiabatic dynamics, where they can reveal dominant motion patterns without prior knowledge.\cite{Peng2021,Zhu2022}  However, because PCs are constructed solely based on structural variance, they may fail to isolate the specific nuclear distortions that control electronic transitions, especially when these are localized or weakly represented in the global variance.\cite{Mai2019} In contrast, the DII-selected internal coordinates retain direct physical interpretability while achieving comparable, or in several cases superior predictive power, highlighting their suitability for mechanistic analysis of nonadiabatic processes.

%%%%%%%%%%%%%%%%%%%%%%%%
\section{Discussion}\label{sec:discussion}
%%%%%%%%%%%%%%%%%%%%%%%%
Across all systems studied here, the information-imbalance workflow provides a consistent strategy to move from high-dimensional trajectory data to a compact, physically interpretable set of decay coordinates. The main practical outcome is that we can rank candidate descriptors by information content and then map those rankings to mechanistic modes, rather than imposing a reaction coordinate \emph{a priori}. This is especially useful for excited-state dynamics, where multiple coupled distortions can coexist and where local CI descriptors are not always straightforward to transfer to dynamical ensembles.

Several system-level trends emerge from the results. First, the method distinguishes localized from collective decay channels in a chemically transparent way. Methaniminium shows a relatively localized dominant coordinate for $S_2/S_1$ but a more collective $S_1/S_0$ channel, while furan displays broad collectivity with coupled puckering and ring-opening coordinates. In L-gln, the key energy-gap signal remains strongly localized (carbonyl stretching/proton-transfer coordinates), whereas in L-pyro the same observables are more distributed and require a broader coordinate set although the dominant modes are still the same carbonyl stretching and proton-transfer coordinates. For molecular motors, the analysis identifies torsion around the central bond as the parent mode governing both non-radiative decay and bright-to-dark conversion.

%The molecular-motor case is, in our view, a particularly novel outcome of this study. 
The molecular-motor case highlights how the workflow can clarify competing mechanistic interpretations in strongly coupled photochemical systems. 
Prior work has debated whether bright-state quenching is primarily associated with pyramidalization or with central-bond torsional/elongation coordinates.\cite{Conyard2012,Wen2023,Kazaryan2011} Here, the DII analysis identifies the central torsional coordinate as the most informative descriptor for both $\Delta E_{1,0}$ and $f^{\mathrm{osc}}_{1,0}$, while the central bond stretch and pyramidalization appear as secondary, coupled responses. 
Thus, the analysis supports a picture in which torsional motion around the central bond acts as the parent coordinate that explains both energy-gap closure and the change of the oscillator strength.
Equally important, this parent mode is recovered even when the initial CM-filtering stage does not directly isolate all dihedral atoms, highlighting the added mechanistic value of the full multi-step II workflow for strongly coupled photochemical pathways.

Second, the analysis reveals a systematic asymmetry between electronic targets: the energy gap is generally easier to explain than the oscillator strength using simple interpretable structural descriptors.  This distinction is physically expected because the oscillator strength contains both an energetic and a transition-dipole moment contribution.
%, \(f^{\mathrm{osc}} \propto \Delta E |\boldsymbol{\mu}_{ij}|^2\)
To disentangle these two effects, we additionally analyzed the transition-dipole contribution alone. This decomposition shows an interesting duality: in the case of L-gln and L-pyro we observe that both the extent to which the internal coordinates describe $f^{\mathrm{osc}}$ and $f^{\mathrm{osc}}\mathrm{(TDM)}$ (as reflected by the value of the DII), as well as the dominant coordinates involved in the description remain consistent. Thus, in these systems, the transition dipole itself is controlled by structural changes that are less localized than the modes that govern the energy gap. 

On the other hand, the molecular motor provides a contrasting case: the same central torsional coordinate governs both \(\Delta E\) and \(f^{\mathrm{osc}}\), and the TDM-only analysis indicates that the bright-to-dark conversion is also organized by the same coupled torsional/bond-stretching motion. However, in this case, the explainability of \(f^{\mathrm{osc}}(\mathrm{TDM})\) is substantially reduced relative to \(f^{\mathrm{osc}}\). This suggests that the stronger structural predictability of \(f^{\mathrm{osc}}\) arises primarily from the explicit contribution of \(\Delta E\). Thus, while the same structural features also influence the TDM contribution, their correlation with \(f^{\mathrm{osc}}(\mathrm{TDM})\) is significantly weaker than with the full oscillator strength.
Overall, these results show that the coordinate controlling energy gap closure is not generally identical to the coordinate controlling optical activity, although the two can coincide in systems where the same structural distortion controls both non-radiative decay and electronic-state character.

%Second, a systematic asymmetry appears between targets: the energy gap is generally easier to explain than oscillator strength with low-order structural descriptors. In multiple systems, $f^{\mathrm{osc}}$ saturates at higher DII values and benefits more from collective representations, indicating that bright/dark behavior often depends on broader coupled motions than the dominant CI-access coordinate. This distinction is important when interpreting photophysics: the coordinate that controls hopping is not necessarily identical to the coordinate that best controls optical activity within the excited state.

This approach is also data-efficient. In particular, the L-pyro case shows that mechanistically meaningful modes can be recovered even when only a limited fraction of trajectories exhibits hopping events. This is relevant for realistic NAD simulations, where computational cost can restrict analysis to a few thousand representative frames (often only a subset of all generated configurations). Even in these data-limited settings, the internal-coordinate ranking remains robust enough to extract dominant channels.

Principal component (PC) descriptors are useful for identifying collective structural distortions in NAD simulations that explain the energy gap. However, they are not universally optimal for predicting electronic observables. In cases where the relevant photochemical process is localized, explicit internal coordinates can provide superior predictive performance.
Furthermore, the most informative PCs are not necessarily those that capture the largest structural variance, and their relevance must therefore be evaluated explicitly. In this sense, PC analysis should be viewed as a diagnostic tool for detecting collective motions that correlate with electronic observables (such as energy gaps or oscillator strengths), rather than as a substitute for physically interpretable internal coordinates.
The II-based approach can also be used to determine the most informative principal components in predicting the energy gap which could then in turn be used to create low-dimensional representations of systems where the nuclear motion is also described quantum mechanically, such as in multiconfigurational time-dependent Hartree (MCTDH) with linear vibronic coupling (LVC) models. This is an extension of the approach done previously in Ref.\citenum{Mai2019} where normal modes were selected based on a linear fit to energy gaps.

In conclusion, the II-based protocol offers a unified and scalable route to identify non-radiative decay modes directly from trajectory ensembles, while retaining mechanistic interpretability across systems ranging from small benchmarks to complex fluorescent materials and molecular motors.

%\deb{COMMENT ON LVC/MCTDH/full QD. Comment on using this for triplets where modes are more collective and it is hard to figure out. Left for future work etc. PCs are not converged. Fine without statistics - analysis often done on a few thousand frames (5-10\%) of all frames. top PCs aren't always PC1/PC2 for eg.}

%%%%%%%%%%%%%%%%%%%%%%%%
%%%%%%%%%%%%%%%%%%%%%%%%

\section*{Data Availability}
The programs and workflow to reproduce the results are available at \url{https://github.com/debarshibanerjee/ML_Conical_Intersections}. The trajectory and full analysis data is available from the corresponding author on reasonable request.

%%%%%%%%%%%%%%%%%%%%%%%%

\begin{acknowledgement}
DB, GDM, and AH acknowledge the funding received by the European Research Council (ERC) under the European Union’s Horizon 2020 research and innovation program (Grant No. 101043272 - HyBOP). The views and opinions expressed are those of the authors only and do not necessarily reflect those of the European Union or the European Research Council Executive Agency. Neither the European Union nor the granting authority can be held responsible for them.
DB, GDM, and AH also acknowledge CINECA for their resources on the cluster Leonardo (from the Convenzione triennale ICTP Anni 2024-2026 project). DB and AH acknowledge the use of AI tools for improving the text and readability of the manuscript.
\end{acknowledgement}

%%%%%%%%%%%%%%%%%%%%%%%%
\bibliography{refs}
%%%%%%%%%%%%%%%%%%%%%%%%

\include{SI}

\end{document}

%% file: figs/flowchart_v2.tex
\begin{figure}[htbp]
\centering
\begin{tikzpicture}[
  node distance = 6mm and 0mm,
  base/.style = {
    rectangle, rounded corners=5pt, thick, align=center, font=\footnotesize
  },
  blueish/.style = {
    base, draw=blue!70!black, fill=blue!6, text width=48mm, minimum height=13mm
  },
  greyish/.style = {
    base, draw=black!60, fill=black!10, text width=48mm, minimum height=13mm
  },
  teal/.style = {
    base, draw=teal!70!black, fill=teal!12, text width=72mm, minimum height=12mm
  },
  dii/.style = {
    base, draw=orange!80!red!70, fill=orange!12!red!4, text width=72mm, minimum height=12mm
  },
  reddish/.style = {
    base, draw=red!70!black, fill=red!12, text width=48mm, minimum height=13mm
  },
  steplabel/.style = {
    font=\sffamily\bfseries, text=black!60, anchor=west
  },
  arr/.style = {
    ->, -{Stealth[length=2.6mm,width=1.9mm]}, line width=1.1pt, color=black!70
  }
]

%% ─── 1. INITIAL DATA AND DESCRIPTORS ───────────────────
\node[greyish] (traj) at (0, 0) {%
  \textbf{NAD trajectory}\\[2pt]
  $\{X^t, \Delta E^t_{l,l-1}\}$};

\node[blueish, below=8mm of traj] (desc) {%
  \textbf{Descriptors}\\[2pt]
  $\mathrm{CM}_{ij}^{S_l}, R^{S_l}$};

%% ─── 2. PIPELINE STEPS ─────────────────────────────────

% Step 1
\node[teal, below=8mm of desc] (step1) {%
  Build $P(\mathrm{CM}_{ij} | S_l)$, $P(\mathrm{CM}_{ij} | S_{l-1})$  
  %for all frames
  };
\node[steplabel] at ([xshift=4mm]step1.east) {Step 1};

% Step 2
\node[teal, below=8mm of step1] (step2) {%
  $\text{JSD}\!\left(P(\mathrm{CM}_{ij}|S_l), P(\mathrm{CM}_{ij}|S_{l-1})\right) > \text{Threshold}$\\[4pt]
  $\downarrow$\\[4pt]
  $\mathrm{CM}_{\mathrm{hotspot}}^{S_l}$};
\node[steplabel] at ([xshift=4mm]step2.east) {Step 2};

% Step 3
\node[dii, below=8mm of step2] (step3) {%
  { $\mathrm{DII}\!\left(\mathrm{CM}_{\mathrm{hotspot}}^{S_l} \rightarrow \Delta E_{l,l-1}\right)$}
  $\rightarrow \mathrm{CM}^{S_l}_\mathrm{f}$ };
\node[steplabel] at ([xshift=4mm]step3.east) {Step 3};

% Step 4
\node[dii, below=8mm of step3] (step4) {%
  {$\mathrm{DII}\!\left(R^{S_l} \rightarrow \mathrm{CM}^{S_l}_\mathrm{f}\right)$}
  $\rightarrow R_{\mathrm{selected}}^{S_l}$};
\node[steplabel] at ([xshift=4mm]step4.east) {Step 4};

% Step 5
\node[dii, below=8mm of step4] (step5) {%
  {$\mathrm{DII}\!\left(R_{\mathrm{selected}}^{S_l} \rightarrow \Delta E_{l,l-1}\right)$}};
\node[steplabel] at ([xshift=4mm]step5.east) {Step 5};

%% ─── 3. FINAL OUTPUT ───────────────────────────────────
\node[reddish, below=8mm of step5] (final) {%
  \textbf{Interpretable Modes}};

%% ─── ARROWS / ROUTING ──────────────────────────────────
\draw[arr] (traj) -- (desc);
\draw[arr] (desc) -- (step1);
\draw[arr] (step1) -- (step2);
\draw[arr] (step2) -- (step3);
\draw[arr] (step3) -- (step4);
\draw[arr] (step4) -- (step5);
\draw[arr] (step5) -- (final);

\end{tikzpicture}
\caption{Workflow for identifying state-relevant internal coordinates from trajectory data. 
Starting from NAD trajectories, Coulomb matrix descriptors are
constructed and compared across electronic states using the Jensen–Shannon divergence
to identify structurally relevant atom pairs (“hotspots”). 
%State-specific Coulomb-matrix distributions are compared using the Jensen--Shannon divergence to identify hotspot atom pairs. 
These are evaluated via DII against the energy gap to construct a filtered descriptor set $\mathrm{CM}_\mathrm{f}$. Internal coordinates are then selected based on their predictive power with respect to $\mathrm{CM}_\mathrm{f}$ and used for final feature selection against $\Delta E_{l,l-1}$. Boxes highlighted in teal are steps using the descriptors and statistical analysis methods, and in orange are the steps where the DII is used.}
\label{fig:protocol}
\end{figure}

%% file: SI.tex
%\documentclass[journal=jacsat,manuscript=article]{achemso}

%\usepackage[version=3]{mhchem} % Formula subscripts using \ce{}
%\usepackage{tablefootnote}
%\usepackage{threeparttable}  

%\author{Debarshi Banerjee}
%\affiliation{International Centre for Theoretical Physics (ICTP), Strada Costiera 11, 34151 Trieste, Italy}
%\alsoaffiliation{Scuola Internazionale Superiore di Studi Avanzati (SISSA), via Bonomea 265, 34136 Trieste, Italy}

%\author{Gonzalo Díaz Mirón}
%\affiliation{International Centre for Theoretical Physics (ICTP), Strada Costiera 11, 34151 Trieste, Italy}

%\author{Alex Rodriguez}
%\email{alejandro.rodriguezgarcia@units.it}
%\affiliation{International Centre for Theoretical Physics (ICTP), Strada Costiera 11, 34151 Trieste, Italy}
%\alsoaffiliation{Dipartimento di Matematica e Geoscienze, University of Trieste, 34127 Trieste, Italy}

%\author{Ali Hassanali}%
%\email{ahassana@ictp.it}
%\affiliation{International Centre for Theoretical Physics (ICTP), Strada Costiera 11, 34151 Trieste, Italy}

%\title{Supplementary Information for: }

%\abbreviations{NAF,DFTB,NAMD}
%keywords{Non-aromatic fluorescence, Density Functional Tight-Binding, Non adiabatic molecular dynamics}

%\SectionNumbersOn

%\begin{document}

\newpage
\begin{suppinfo}
\setcounter{table}{0}
\renewcommand{\thetable}{S\arabic{table}}
\newcounter{SItab}
\renewcommand{\theSItab}{S\arabic{SItab}}
\setcounter{figure}{0}
\renewcommand{\thefigure}{S\arabic{figure}}
\newcounter{SIfig}
\renewcommand{\theSIfig}{S\arabic{SIfig}}

\begin{figure}[H]
  \centering
  \includegraphics[width=1.0\linewidth]{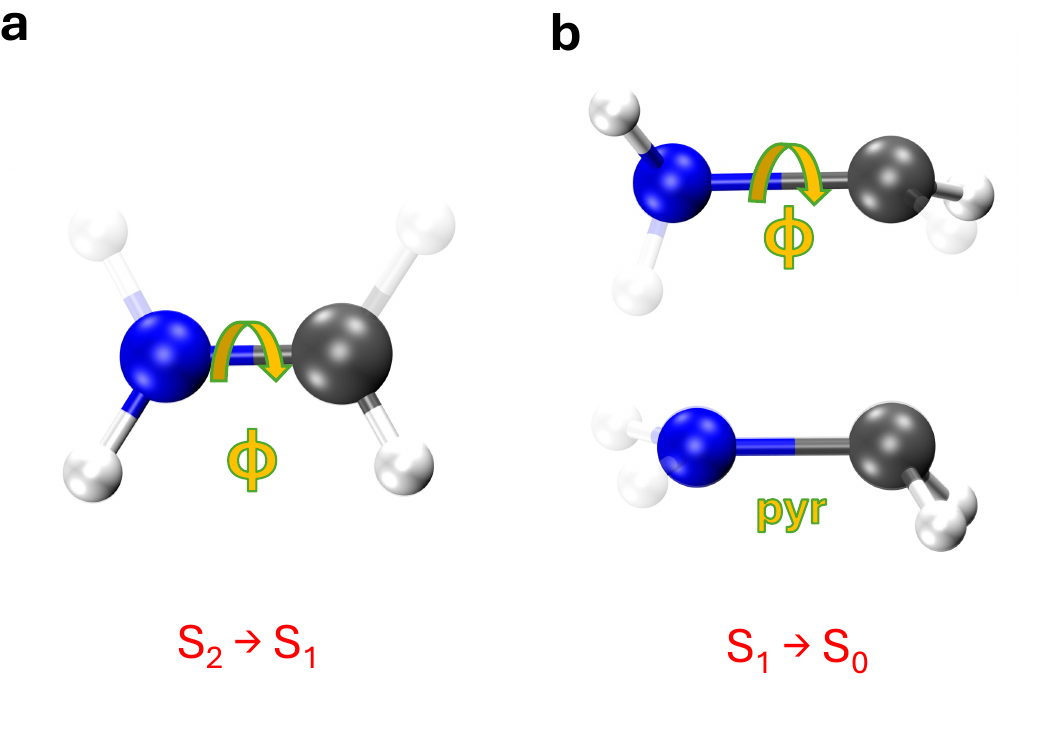}
  \caption{Methaniminium Cation:
Additional internal-coordinate modes for the methaniminium cation: (a) Mode relevant for the S$_2 \rightarrow$ S$_1$ transition which involves a torsion around the \ce{C-N} bond. (b) Modes relevant for the S$_1 \rightarrow$ S$_0$ transition, which involves a torsion around the \ce{C-N} bond as well as the pyramidalization of the \ce{CH2} group.
  }
  \label{fig_SI:formal_extra_modes}
\end{figure}

%\begin{figure}[H]
%  \centering
%  \includegraphics[width=1.0\linewidth]{figs/Fig-SI_MC_dii-%pc_vs_features.pdf}
%  \caption{Methaniminium Cation PC}
%  \label{fig_SI:formal_pc_dii}
%\end{figure}

\begin{figure}[H]
  \centering
  \includegraphics[width=1.0\linewidth]{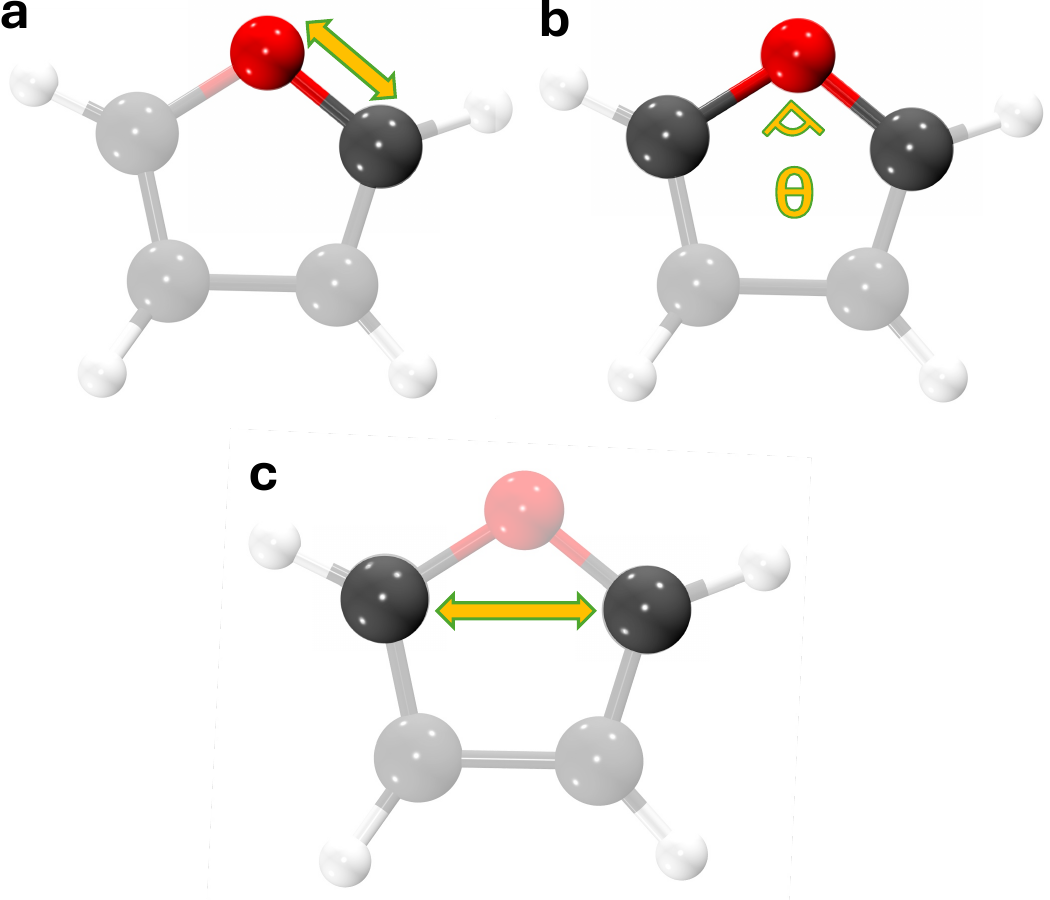}
  \caption{Furan:
Additional internal-coordinate modes for Furan that explain the S$_1 \rightarrow$ S$_0$ non-radiative relaxation: (a) \ce{C-O} stretch. (b) \ce{C-O-C} angle. (c) \ce{C-C} separation that corresponds to the ring opening mode.
  }
  \label{fig_SI:furan_extra_modes}
\end{figure}

\begin{figure}[H]
  \centering
  \includegraphics[width=1.0\linewidth]{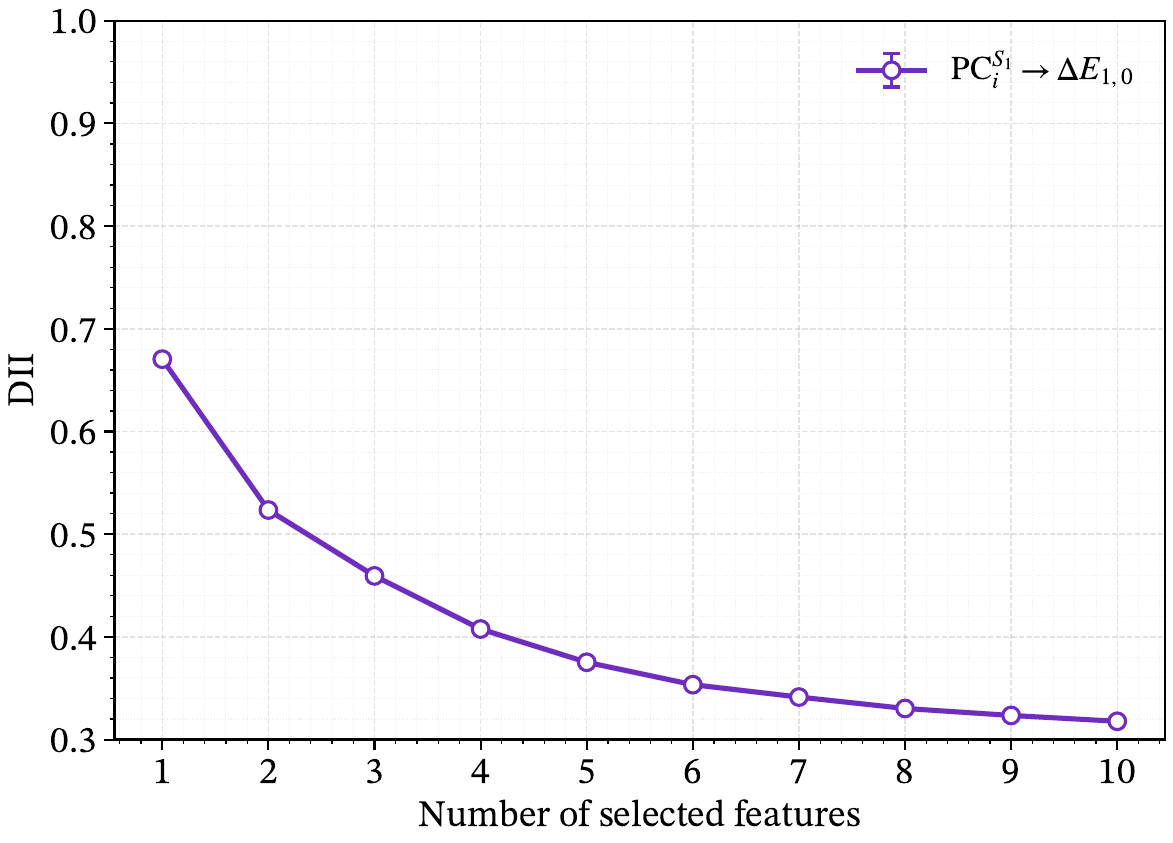}
  \caption{
  Furan: DII between the leading PCs and energy gap ($\Delta E_{1,0}$) as a function of the number of selected features.\\
  The inclusion of PCs continues to lower the DII to $\sim 0.32$. This indicates that, even when a select number of chemically intuitive coordinates already captures part of the dynamics, additional predictive information is contained in more delocalized collective modes. Such behavior reflects the inherently high-dimensional and strongly coupled nature of nonadiabatic dynamics, where multiple nuclear degrees of freedom contribute simultaneously to the evolution toward conical intersections.
  }
  \label{fig_SI:furan_pc_dii}
\end{figure}
\clearpage

\begin{figure}[H]
  \centering
  \includegraphics[width=1.0\linewidth]{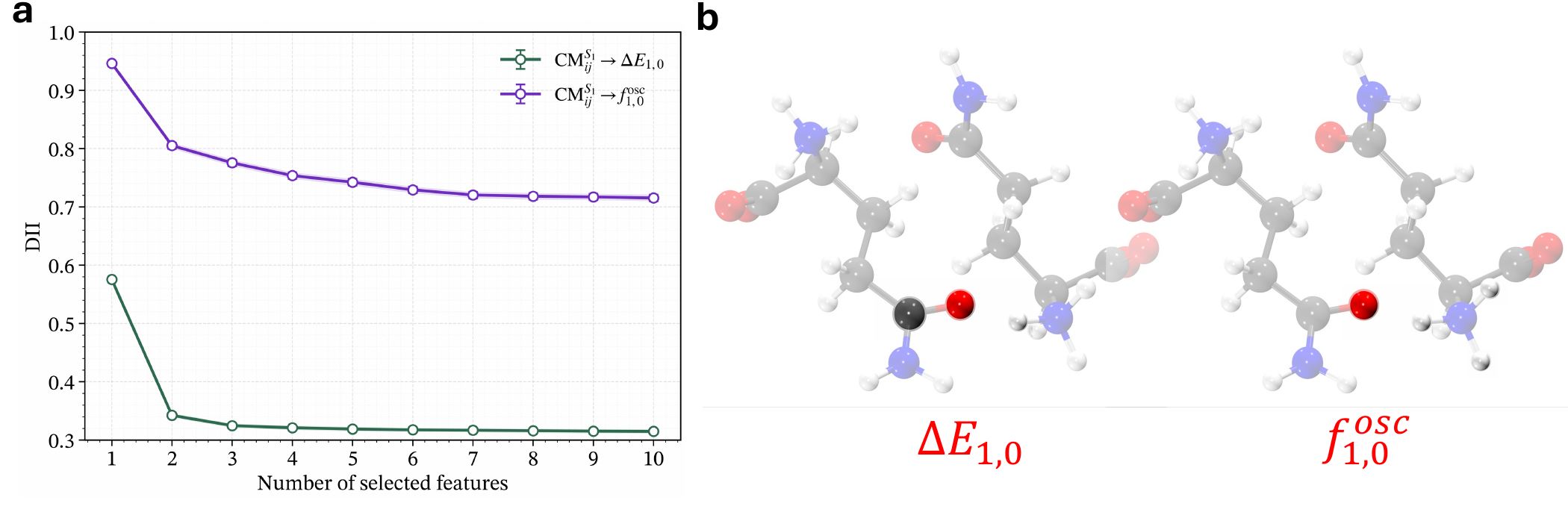}
  \caption{L-gln:
(a) DII between Coulomb-matrix descriptors and energy gap/oscillator strength as a function of the number of selected features. Purple corresponds to the $f^{\mathrm{osc}}_{1,0}$ and green corresponds to the $\Delta E_{1,0}$ respectively.
(b) Most informative atom pairs (“hotspots”) identified from Coulomb matrix analysis are highlighted. The left and right parts of the panel corresponds to the $\Delta E_{1,0}$ and $f^{\mathrm{osc}}_{1,0}$ respectively. 
  }
  \label{fig_SI:gln_cmat}
\end{figure}
\clearpage

\begin{figure}[H]
  \centering
  \includegraphics[width=1.0\linewidth]{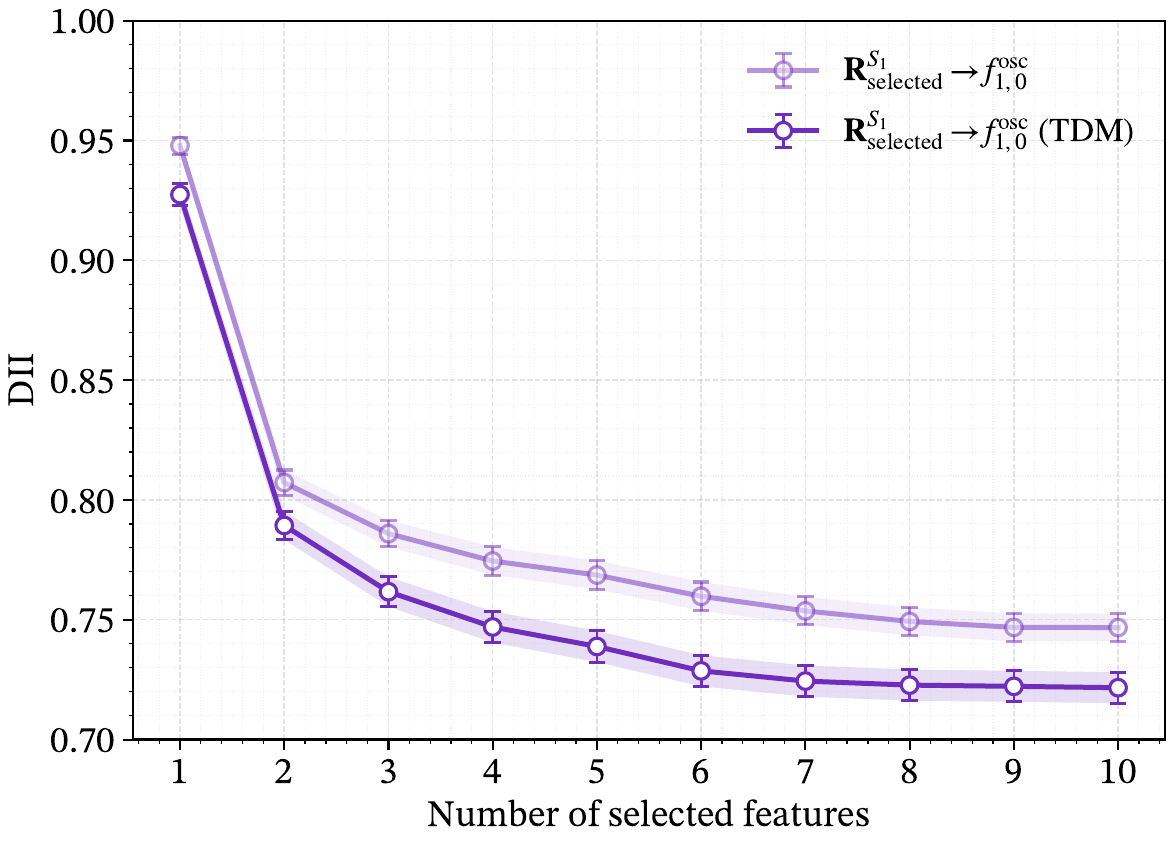}
  \caption{L-gln: DII between internal coordinates and the oscillator strength ($f^{\mathrm{osc}}_{1,0}$) and the part of the oscillator strength corresponding to the Transition Dipole Moment only ($f^{\mathrm{osc}}_{1,0}\mathrm{(TDM)}$) as a function of the number of selected features in faint purple and more bold purple respectively.}
  \label{fig_SI:gln_fosc_tdm}
\end{figure}
\clearpage

\begin{figure}[H]
  \centering
  \includegraphics[width=1.0\linewidth]{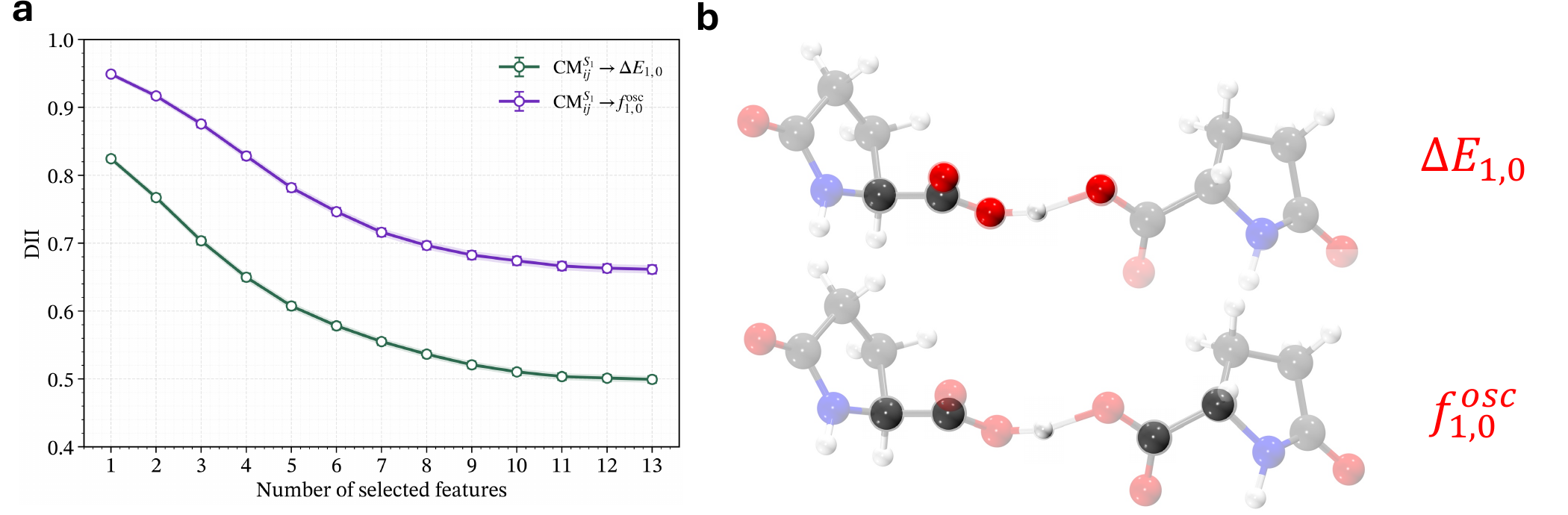}
  \caption{L-pyro:
(a) DII between Coulomb-matrix descriptors and energy gap/oscillator strength as a function of the number of selected features. Purple corresponds to the $f^{\mathrm{osc}}_{1,0}$ and green corresponds to the $\Delta E_{1,0}$ respectively.
(b) Most informative atom pairs (“hotspots”) identified from Coulomb matrix analysis are highlighted. The top and bottom parts of the panel corresponds to the $\Delta E_{1,0}$ and $f^{\mathrm{osc}}_{1,0}$ respectively. 
  }
  \label{fig_SI:pyro_cmat}
\end{figure}
\clearpage

\begin{figure}[H]
  \centering
  \includegraphics[width=1.0\linewidth]{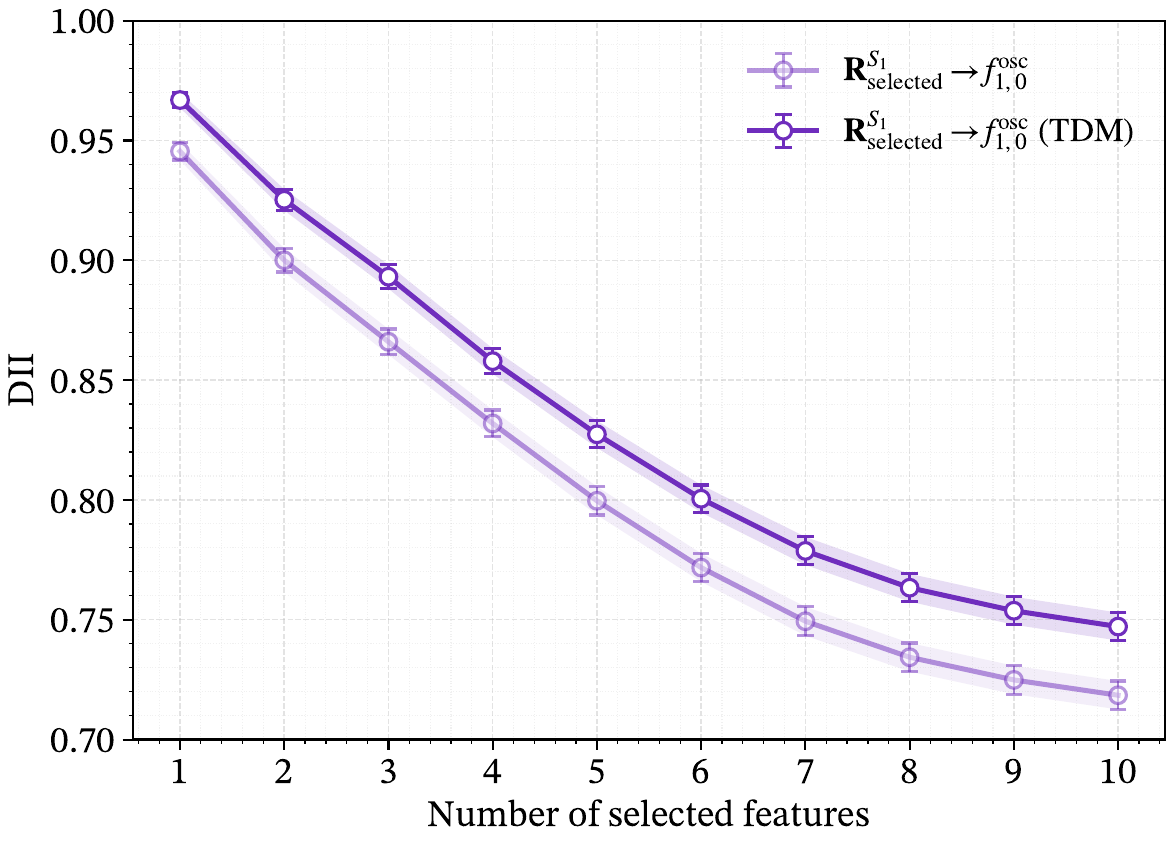}
  \caption{L-pyro: DII between internal coordinates and the oscillator strength ($f^{\mathrm{osc}}_{1,0}$) and the part of the oscillator strength corresponding to the Transition Dipole Moment only ($f^{\mathrm{osc}}_{1,0}\mathrm{(TDM)}$) as a function of the number of selected features in faint purple and more bold purple respectively.}
  \label{fig_SI:pyro_fosc_tdm}
\end{figure}

%\begin{figure}[H]
%  \centering
%  \includegraphics[width=1.0\linewidth]{figs/Fig-SI_Gln_dii-pc_vs_features.pdf}
%  \caption{L-gln PC}
%  \label{fig_SI:gln_pc_dii}
%\end{figure}

\begin{figure}[H]
  \centering
  \includegraphics[width=1.0\linewidth]{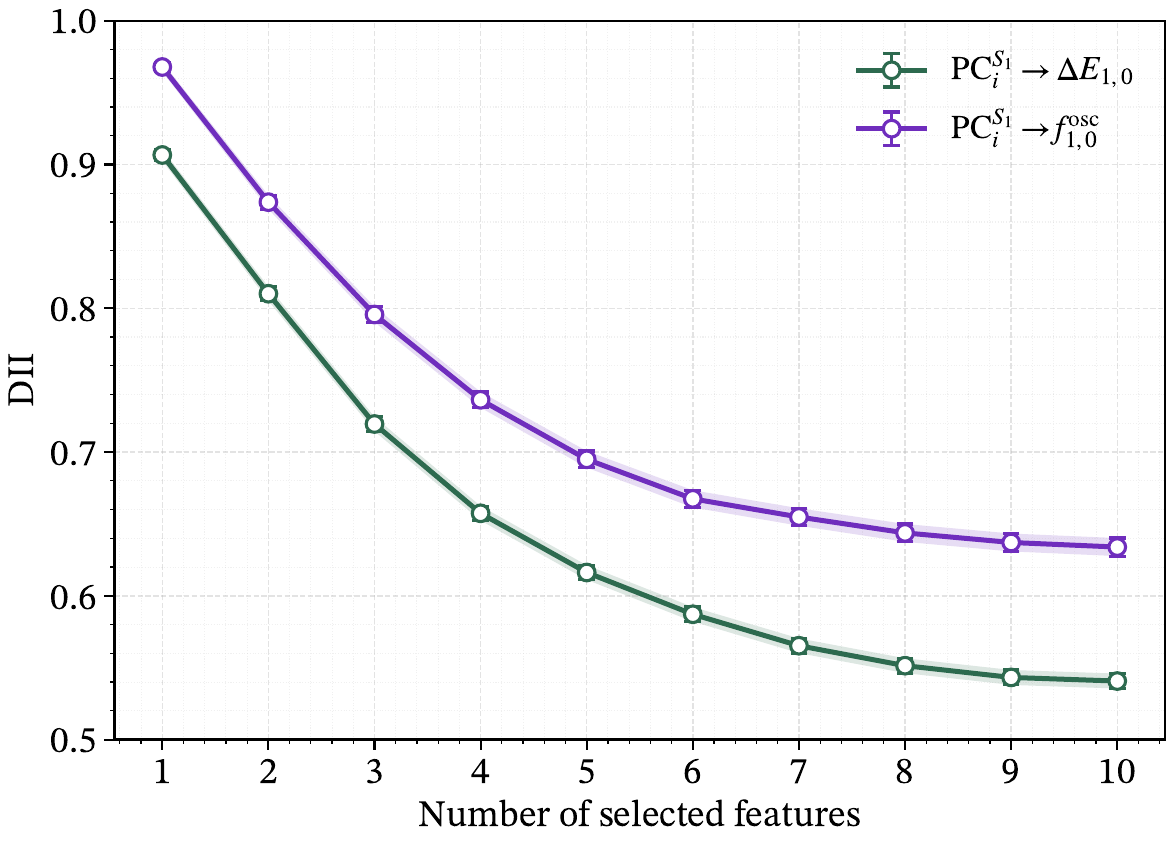}
  \caption{
  L-pyro: DII between the leading PCs and energy gap/oscillator strength as a function of the number of selected features. Purple corresponds to the $f^{\mathrm{osc}}_{1,0}$ and green corresponds to the $\Delta E_{1,0}$ respectively.
  }
  \label{fig_SI:pyro_pc_dii}
\end{figure}

\begin{figure}[H]
  \centering
  \includegraphics[width=1.0\linewidth]{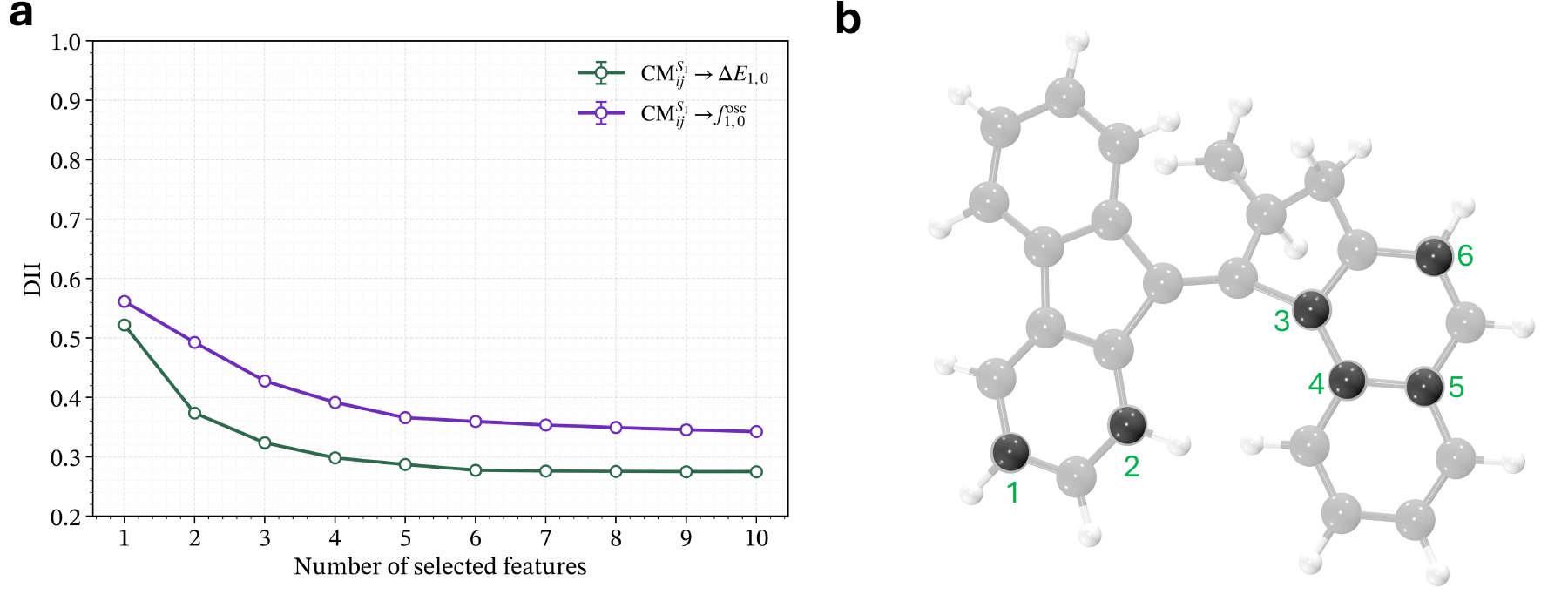}
  \caption{Molecular Motor:
(a) DII between Coulomb-matrix descriptors and energy gap/oscillator strength as a function of the number of selected features. Purple corresponds to the $f^{\mathrm{osc}}_{1,0}$, red corresponds to $f^{\mathrm{osc}}_{1,0}\mathrm{(TDM)}$, and green corresponds to the $\Delta E_{1,0}$ respectively.
(b) Most informative atom pairs (“hotspots”) identified from Coulomb matrix analysis are highlighted. The atoms involved in the hotspots for both $\Delta E_{1,0}$ and $f^{\mathrm{osc}}_{1,0}$ are highlighted together and numbered to distinguish the different C atoms. 
}
  \label{fig_SI:motors_cmat}
\end{figure}

\begin{figure}[H]
  \centering
  \includegraphics[width=1.0\linewidth]{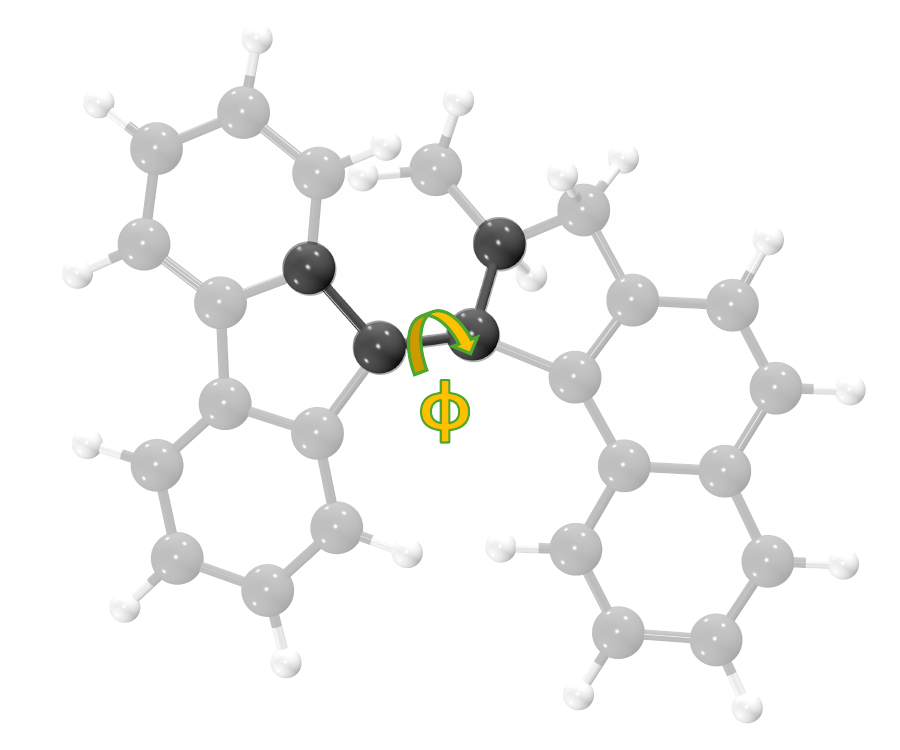}
  \caption{Molecular Motor:
Additional internal-coordinate mode involving a secondary torsion around the central dihedral for the molecular motor system that explain the S$_1 \rightarrow$ S$_0$ energy gap ($\Delta E_{1,0}$) as well as the oscillator strength ($f^{\mathrm{osc}}_{1,0}$).
  }
  \label{fig_SI:motors_extra_modes}
\end{figure}

\begin{figure}[H]
  \centering
  \includegraphics[width=1.0\linewidth]{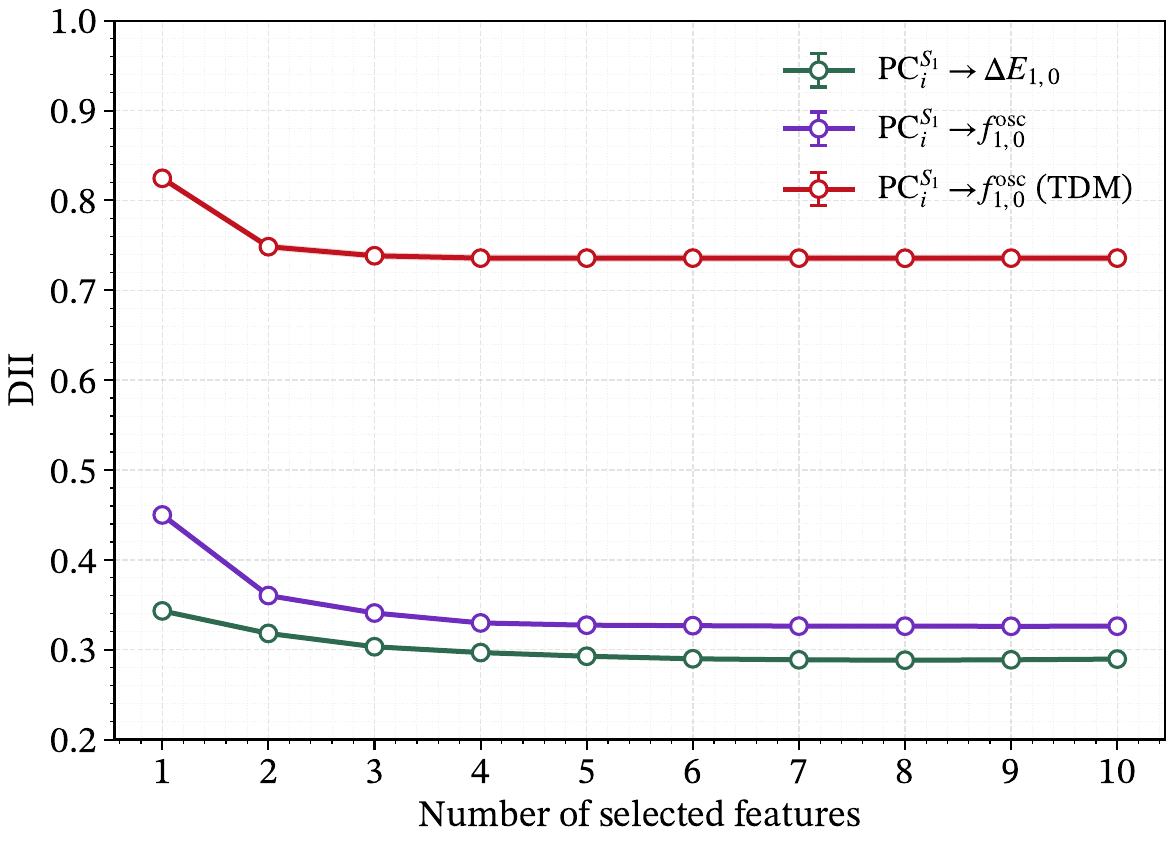}
  \caption{
  Molecular Motor: DII between the leading PCs and energy gap/oscillator strength as a function of the number of selected features. Purple corresponds to the $f^{\mathrm{osc}}_{1,0}$ and green corresponds to the $\Delta E_{1,0}$ respectively.
  }
  \label{fig_SI:motors_pc_dii}
\end{figure}

\end{suppinfo}

%\bibliography{bibio}

%\end{document}